\documentclass[journal,article,submit,moreauthors]{mdpi} 
\UseRawInputEncoding
\firstpage{1} 
\hreflink{https://doi.org/} % If needed use \linebreak

\usepackage{wrapfig}
\usepackage[pscoord]{eso-pic}
\usepackage[fulladjust]{marginnote}
\usepackage{titletoc}   

\titlecontents{section}[0.6em]{\vspace{.25\baselineskip}}
             {\thecontentslabel\quad}{}
             {\hspace{.5em}\titlerule*[10pt]{$\cdot$}\contentspage}
\titlecontents{subsection}[0.6em]{\vspace{.25\baselineskip}}
             {\thecontentslabel\quad}{}
             {\hspace{.5em}\titlerule*[10pt]{$\cdot$}\contentspage}
 \titlecontents{subsubsection}[0.6em]{\vspace{.25\baselineskip}}
             {\thecontentslabel\quad}{}
             {\hspace{.5em}\titlerule*[10pt]{$\cdot$}\contentspage}            
\reversemarginpar

\definecolor{Gray}{gray}{.25}
\Title{\boldmath Probability Thermodynamics and Probability Quantum Field } 

\Author{Ping Zhang $^{1,\dagger}$, Wen-Du Li $^{2,\ddagger}$ , Tong Liu $^{3,\S}$ , and Wu-Sheng Dai $^{4,\textbar\textbar}$}

\address{%
$^{1}$ \quad School of Finance, Capital University of Economics and Business, Beijing 100070, P. R. China\\
$^{2}$ \quad College of Physics and Materials Science, Tianjin Normal University, Tianjin 300387, PR China\\
$^{3}$ \quad School of Architecture, Tianjin University, Tianjin 300072, P. R. China\\
$^{4}$ \quad Department of Physics, Tianjin University, Tianjin 300350, P.R. China
}

\firstnote{zhangping@cueb.edu.cn}
\secondnote{liwendu@tjnu.edu.cn}
\thirdnote{liutong@tju.edu.cn} 
\fourthnote{daiwusheng@tju.edu.cn.}
\abstract{We introduce probability thermodynamics and probability quantum fields. By
probability we mean that there is an unknown operator, physical or
nonphysical, whose eigenvalues obey a certain statistical distribution.
Eigenvalue spectra define spectral functions. Various thermodynamic quantities
in thermodynamics and effective actions in quantum field theory are all
spectral functions. In the scheme, eigenvalues obey a probability
distribution, so a probability distribution determines a family of spectral
functions in thermodynamics and quantum field theory. This leads to
probability thermodynamics and probability quantum fields determined by a
probability distribution. In constructing spectral functions, we encounter a
problem. The conventional definition of spectral functions applies only to
lower bounded spectra. In our scheme, however, there are two types of spectra:
lower bounded spectra, corresponding to the probability distribution with
nonnegative random variables, and the lower unbounded spectra, corresponding
to probability distributions with negative random variables. To take the lower
unbounded spectra into account, we generalize the definition of spectral
functions by analytical continuation. In some cases, we encounter divergences.
We remove the divergence by a renormalization procedure. Moreover, in virtue
of spectral theory in physics, we generalize some concepts in probability
theory. For example, the moment-generating function in probability theory does
not always exist. We redefine the moment-generating function as the
generalized heat kernel introduced in this paper, which makes the concept
definable when the definition in probability theory fails. As examples, we
construct examples corresponding to some probability distributions.
Thermodynamic quantities, vacuum amplitudes, one-loop effective actions, and
vacuum energies for various probability distributions are presented.}

\begin{document}
\nolinenumbers

\setcounter{tocdepth}{2}
\begin{sloppypar}
\tableofcontents
\end{sloppypar}

%%%%%%%%%%%%%%%%%%%%%%%%%%%%%%%%%%%%%%%%%%

% now start line numbers
%\linenumbers

% the * after section prevents numbering
%\section*{Introduction}
 
\section{Introduction}

A physical system is described by an operator. The eigenvalue of the operator
contains only partial information about the system. This is the reason why one
\textit{cannot} hear the shape of a drum \cite{kac1966can}. Nevertheless, the
information embedded in eigenvalues, though partial, is of special importance.
In physics, for example, in thermodynamics, various thermodynamic quantities
are determined by eigenvalues, and in quantum field theory the vacuum
amplitude, the effective action, and the vacuum energy are determined by
eigenvalues \cite{vassilevich2003heat}. In mathematics, a famous example of
spectral geometry is given by Kac that one can extract topological
information, the Euler characteristics, from eigenvalues through the heat
kernel or the spectral counting function \cite{kac1966can,dai2009number}.
Quantities, such as thermodynamic quantities, effective actions, and spectral
counting functions, are all spectral functions. The spectral function is
determined by a spectrum $\left\{  \lambda_{n}\right\}  $, and, on the other
hand, all information of the spectrum is embedded in the spectral function
\cite{zhou2018calculating}. Once an eigenvalue spectrum is given, even if the
operator is unknown, we can also construct various spectral functions.

In the scheme, without knowing the operator, we present an eigenvalue spectrum
to construct spectral functions. The eigenvalue spectrum $\left\{  \lambda
_{n}\right\}  $ is supposed to obey a certain probability distribution.
Therefore, spectral functions constructed in this way are determined by the
probability distribution. They are probability spectral functions.
Technologically, in the scheme, the state density of an eigenvalue spectrum is
chosen as a probability density function, or equivalently, the spectral
counting function of an eigenvalue spectrum is chosen as a cumulative
probability function. Constructing the thermodynamic quality with a
probability eigenvalue spectrum, we arrive at probability thermodynamics;
constructing the vacuum amplitude and the effective action, etc., with a
probability eigenvalue spectrum, we arrive at a probability quantum field.

Our starting point is the heat kernel. The local heat kernel $K\left(
t;\mathbf{r},\mathbf{r}^{\prime}\right)  $ of an operator $D$ is the Green
function of the initial-value problem of the heat-type equation: $\left(
\partial_{t}+D\right)  K\left(  t;\mathbf{r},\mathbf{r}^{\prime}\right)  =0$
with $K\left(  0;\mathbf{r},\mathbf{r}^{\prime}\right)  =\delta\left(
\mathbf{r}-\mathbf{r}^{\prime}\right)  $ \cite{vassilevich2003heat}. The
global heat kernel is the trace of the local heat kernel $K\left(
t;\mathbf{r},\mathbf{r}^{\prime}\right)  $:%
\begin{equation}
K\left(  t\right)  =\sum_{n=0}^{\infty}e^{-\lambda_{n}t}. \label{trdht}%
\end{equation}
Here $\{\lambda_{n}\}$ is the eigenvalue spectrum of the operator $D$,
determined by the eigenequation $D\phi_{n}=\lambda_{n}\phi_{n}$. In the
conventional definition, the lowest eigenvalue must be finite. In
thermodynamics, the heat kernel is the partition function; in quantum field
theory, the heat kernel is the vacuum amplitude. From the heat kernel, we can
obtain all thermodynamic quantities in thermodynamics and effective actions
and vacuum energies in quantum field theory.

In probability spectral functions, we encounter unbounded spectra. This is
because there are two types of probability distributions: one includes only
nonnegative random variables, corresponding to lower bounded spectra, and the
other includes negative random variables, corresponding to lower unbounded
spectra. The definition of the heat kernel, Eq. (\ref{trdht}), applies only to
lower bounded spectra $\left\{  \lambda_{0},\lambda_{1},\lambda_{2}%
,\cdots\right\}  $ which has a finite lowest eigenvalue $\lambda_{0}$. For
lower unbounded spectra, the lowest eigenvalue tends to negative infinity and
the sum in Eq. (\ref{trdht}) often diverges, so we need to generalize the
conventional definition of spectral functions.

As mentioned above, for a lower unbounded spectrum, there is no lowest
eigenvalue, i.e., the lowest eigenvalues $\lambda_{0}\rightarrow-\infty$, and
the sum in the definition (\ref{trdht}) often diverges. In this paper, we
first generalize the conventional definition of the heat kernel so as to make
the definition of heat kernel apply to lower unbounded spectra. Then, we
obtain other spectral functions from the heat kernel by the relation between
spectral functions \cite{vassilevich2003heat,dai2009number,dai2010approach}.

It should be emphasized that the eigenvalue spectrum of a real physical system
must be lower bounded, or else the world would be unstable. The system
considered here, however, is a fictional physical system. Their eigenvalue
spectra are not from a physical operator but are required to satisfy a
probability distribution artificially. The eigenvalue spectrum corresponding
to a probability distribution with negative random variables leads to lower
unbounded eigenvalue spectra. Therefore, unsurprisingly, some thermodynamic
qualities may not be positive. That is, the operator corresponding to
probability distributions is sometimes not a physically real operator.

Though the eigenvalue spectrum corresponding to probability distributions is
sometimes not physically real, the probability eigenvalue spectrum has good
behavior. This is because the probability distribution has good behavior, such
as integrability. Therefore, for example, the vacuum energy though always
diverges in quantum field theory, it converges for probability eigenvalue spectra.

The scheme suggested in the present paper also generalizes some concepts in
probability theory. Take the moment-generating function as an example. In
probability theory, the moment-generating function does not always exist. In
this paper, the generalized heat kernel is used to serve as a generalized
moment-generating function. In many cases, the new definition is definable
when the definition of the moment-generating function in probability theory fails.

When constructing quantum fields, in some cases, we encounter divergences. We
remove the divergence by a renormalization procedure.

As examples, we consider some probability distributions, including
distributions without and with negative random variables. The corresponding
thermodynamics and quantum fields are presented.

As a byproduct, we have give a novel derivation for the probability density
function of the intermediate distribution proposed in Ref.
\cite{liu2012intermediate}.

In section \ref{Genspectralfunction}, a scheme for generalizing spectral
functions to lower unbounded spectra is suggested. In section
\ref{Probabilityspectralfunction}, the probability spectral function is
introduced. In sections \ref{Thermodynamics} and \ref{QFT}, probability
thermodynamics and probability quantum field theory are constructed. In
section \ref{various}, we illustrate the construction of probability
thermodynamics and probability quantum field theory through examples. The
conclusions and outlook are provided in section \ref{Conclusions}. In
appendices \ref{nonnegative} and \ref{negative}, we list probability
thermodynamics and probability quantum field theory for various distributions.
In appendix \ref{Inter}, we give an alternative derivation of the distribution
function for the intermediate distribution.

\section{Generalized spectral function \label{Genspectralfunction}}

Unlike the spectral function in physics, which is determined by the eigenvalue
spectrum $\left\{  \lambda_{n}\right\}  $ of a lower bounded Hermitian
operator $D$, the eigenvalue spectrum we are interested in is given by a
probability distribution, which may be lower unbounded. For a lower bounded
spectrum, there is a finite lowest eigenvalue $\lambda_{\min}$, but for a
lower unbounded spectrum, the lowest eigenvalue tends to negative infinity:
$\lambda_{\min}\rightarrow-\infty$. The conventional definition of spectral
functions is valid for lower-bounded spectra. One aim of the present paper is
to generalize the definition of spectral functions to lower unbounded spectra.
There are also discussions on the generalized spectral functions in literature
\cite{fursaev2011operators,birrell1984quantum}.

In this section, we generalize the definition of some essential spectral
functions, the heat kernel, the spectral counting function, and the spectral
zeta function, to lower unbounded spectra. These spectral functions are
essentially important in spectral geometry, thermodynamics, and quantum field theory.

\subsection{Generalized heat kernel}

\subsubsection{Generalized heat kernel: Wick rotation \label{defintion}}

In this section, we generalize the conventional definition of the global heat
kernel to make it valid for both bounded and unbounded spectra.

The conventional definition of the global heat kernel for a lower bounded
spectrum $\{\lambda_{0},\lambda_{1},\lambda_{2},\cdots\}$, by Eq.
(\ref{trdht}), is $K\left(  t\right)  =\sum_{n=0}^{\infty}e^{-\lambda_{n}t}$.
For lower unbounded spectra, $\{-\infty,\cdots,\lambda_{-1},$ $\lambda
_{0},\lambda_{1},\cdots\}$, a natural and naive generalization of the heat
kernel, following the original intention of the conventional definition of the
heat kernel, which is a sum of $e^{-\lambda_{n}t}$ over all eigenstates, is
\begin{equation}
K\left(  t\right)  =\sum_{n=-\infty}^{\infty}e^{-\lambda_{n}t}. \label{Kgen}%
\end{equation}
This is, however, ill-defined because the existence of the divergent lowest
eigenvalue $\lambda_{-\infty}=-\infty$ brings a divergent term $e^{-\lambda
_{-\infty}t}$ in the summation.

In order to seek a generalized heat kernel for taking lower unbounded spectra
into account, we employ the Wick rotation.

Wick rotation performs the replacement $t=-i\tau$. (In physics, the Wick
rotation usually takes $t=i\tau$. The reason why we take $t=-i\tau$ is that we
want to apply the result to probability theory.) After the Wick rotation , the
definition (\ref{Kgen}) becomes%
\begin{equation}
K^{WR}\left(  \tau\right)  =\sum_{n=-\infty}^{\infty}e^{i\lambda_{n}\tau},
\label{KtWick}%
\end{equation}
where $K^{WR}\left(  \tau\right)  $ denotes the Wick-rotated heat kernel.

After the Wick rotation, we may obtain a finite result $K^{WR}\left(
\tau\right)  $ for lower unbounded spectra $\{-\infty,\cdots,\lambda
_{-1},\lambda_{0},\lambda_{1},\cdots\}$. Nevertheless, the Wick-rotated heat
kernel $K^{WR}\left(  \tau\right)  $ is, of course, not the heat kernel we
want. To obtain a heat kernel rather than a Wick-rotated heat kernel, after
working out the sum in Eq. (\ref{KtWick}), we perform an inverse Wick rotation
$\tau=it$. After the inverse Wick rotation, the Wick-rotated heat kernel
returns back to the heat kernel we want. The heat kernel obtained this way is finite.

In short, the generalized heat kernel for a lower unbounded spectrum
$\{-\infty,\cdots,\lambda_{-1},\lambda_{0},$ $\lambda_{1},\cdots\}$ is obtained
by the following procedure:%
\begin{equation}
K\left(  t\right)  =\left.  \sum_{n=-\infty}^{\infty}e^{i\lambda_{n}\tau
}\right\vert _{\tau=it}=K^{WR}\left(  it\right)  . \label{genKt}%
\end{equation}
Such a procedure is essentially an analytic continuation, which allows us to
achieve a finite result of a divergent series.

In this procedure, there are two steps to achieve the well-defined generalized
heat kernel for lower unbounded spectra, Eq. (\ref{genKt}). In the first step,
instead of the divergent summation $\sum_{n=-\infty}^{\infty}e^{-\lambda_{n}%
t}$\ in the naive generalization of heat kernel (\ref{Kgen}), we turn to
calculate a convergent Wick-rotated summation $\sum_{n=-\infty}^{\infty
}e^{i\lambda_{n}\tau}$. In the second step, we perform an inverse Wick
rotation to the already worked-out result of the Wick-rotated sum. Clearly,
the key step in this procedure is to convert the divergent sum $\sum
_{n=-\infty}^{\infty}e^{-\lambda_{n}\tau}$ to a convergent sum $\sum
_{n=-\infty}^{\infty}e^{i\lambda_{n}\tau}$ by the Wick rotation.

Now, we arrive at a definition, (\ref{genKt}), defining a generalized heat
kernel, valid for lower unbounded spectra.

\subsubsection{Generalized heat kernel: two-sided Laplace transformation and
Fourier transformation \label{intdefintion}}

In the above, the generalized heat kernel is expressed in a summation form. By
the state density
\begin{equation}
\rho\left(  \lambda\right)  =\sum_{\left\{  \lambda_{n}\right\}  }%
\delta\left(  \lambda-\lambda_{n}\right)  , \label{statedensity}%
\end{equation}
we can rewrite the generalized heat kernel in an integral form. From the
integral representation, we can see how the Wick rotation treatment works more intuitively.

\textit{Conventional heat kernel.} First, let us consider the conventional
definition of heat kernels, Eq. (\ref{trdht}), which is only valid for lower
bounded spectra.

By the state density (\ref{statedensity}), we can convert the sum in the
conventional definition (\ref{trdht}) into an integral,%
\begin{equation}
K\left(  t\right)  =\int_{0}^{\infty}d\lambda\rho\left(  \lambda\right)
e^{-\lambda t}. \label{conKtint}%
\end{equation}
The equivalence between the summation form (\ref{trdht}) and the integral form
(\ref{conKtint}) can be checked directly by substituting the state density
(\ref{statedensity}) into Eq. (\ref{conKtint}):\newline$K\left(  t\right)
=\sum_{\left\{  \lambda_{n}\right\}  }\int_{0}^{\infty}d\lambda\delta\left(
\lambda-\lambda_{n}\right)  e^{-\lambda t}=\sum_{n=0}^{\infty}e^{-\lambda
_{n}t}$.

For lower bounded spectra, the integral representation (\ref{conKtint}) of
heat kernels shows that the relation between the heat kernel and the state
density is a Laplace transformation.

\textit{Naive generalized heat kernel.} Next, let us see the naive
generalization of heat kernels, Eq. (\ref{Kgen}).

Converting the sum in Eq. (\ref{Kgen}) into an integral gives%
\begin{equation}
K\left(  t\right)  =\int_{-\infty}^{\infty}d\lambda\rho\left(  \lambda\right)
e^{-\lambda t}. \label{Ktint}%
\end{equation}
Notice that for lower unbounded spectra, $\left\{  \lambda_{n}\right\}  $ runs
from $-\infty$ to $+\infty$.

For lower unbounded spectra, by inspection of Eq. (\ref{Ktint}), we can see
that the relation between the heat kernel and the state density is a two-sided
Laplace transformation (also known as the bilateral Laplace transformation)
\cite{lepage2012complex} rather than a Laplace transformation. The relation
between a two-sided Laplace transformation and Laplace transformation is
$\mathcal{BL}\left[  f\left(  \lambda\right)  ;t\right]  =\mathcal{L}\left[
f\left(  \lambda\right)  ;t\right]  +\mathcal{L}\left[  f\left(
-\lambda\right)  ;-t\right]  $, where $\mathcal{BL}\left[  f\left(
\lambda\right)  ;t\right]  $ denotes the two-sided Laplace transformation and
$\mathcal{L}\left[  f\left(  \lambda\right)  ;t\right]  $ denotes the Laplace transformation.

It is worth pointing out here that the integral form of the naive generalized
heat kernel (\ref{Ktint}) is also valid for a certain kind of lower unbounded
spectra, though the sum form of the naive generalized heat kernel, Eq.
(\ref{Kgen}), is invalid for all kinds of lower unbounded spectra. This is
because in the summation form of the generalized heat kernel, Eq.
(\ref{Kgen}), the term $e^{-\lambda_{-\infty}t}\ $diverges when $\lambda
_{-\infty}=-\infty$; while, in the integral form, Eq. (\ref{Ktint}), the
validity of the definition relies on the integrability of the integral,
$\int_{-\infty}^{\infty}d\lambda\rho\left(  \lambda\right)  e^{-\lambda t}$.
If the integral is integrable, then Eq. (\ref{Ktint}) can serve as a
definition of heat kernels for lower unbounded spectra. The integrability
condition, clearly, is that the state density, $\rho\left(  \lambda\right)  $,
must attenuate rapidly enough. In other words, the definition (\ref{Ktint}) is
not valid for all kinds of lower unbounded spectra; it is only valid for
$\rho\left(  \lambda\right)  $ decreasing faster than $e^{\lambda t}$.

\textit{Wick-rotated heat kernel. }Now, let us see the Wick rotated heat
kernel, Eq. (\ref{KtWick}). Converting Eq. (\ref{KtWick}) into an integral
gives%
\begin{equation}
K^{WR}\left(  \tau\right)  =\int_{-\infty}^{\infty}\rho\left(  \lambda\right)
e^{i\lambda\tau}d\lambda. \label{KtWR}%
\end{equation}
The Wick rotated heat kernel $K^{WR}\left(  \tau\right)  $ is obviously a
Fourier transformation of the state density $\rho\left(  \lambda\right)  $.

Now, we can explain why the Wick rotated treatment is needed for lower
unbounded spectra. Without the Wick rotation, the integral form of the naive
generalization of the heat kernel, Eq. (\ref{Ktint}), is a two-sided Laplace
transformation, and the integrability condition is that the state density
$\rho\left(  \lambda\right)  $ decreases faster than $e^{\lambda t}$.
Nevertheless, the Wick rotated heat kernel, as shown in Eq. (\ref{KtWR}), is a
Fourier transformation of the state density $\rho\left(  \lambda\right)  $.
The integrability condition then becomes that $\rho\left(  \lambda\right)  $
is absolutely integrable, which is easier to be satisfied than that in the
two-sided Laplace transformation.

\textit{Generalized heat kernel. }The generalized heat kernel can be finally
obtained by performing an inverse Wick rotation to the Wick rotated heat
kernel $K^{WR}\left(  \tau\right)  $. From Eq. (\ref{KtWR}), we can see that
the state density $\rho\left(  \lambda\right)  $ is an inverse Fourier
transformation of the Wick rotated heat kernel $K^{WR}\left(  \tau\right)  $,
i.e.,
\begin{equation}
\rho\left(  \lambda\right)  =\frac{1}{2\pi}\int_{-\infty}^{\infty}%
K^{WR}\left(  \tau\right)  e^{-i\lambda\tau}d\tau.
\end{equation}
Substituting into Eq. (\ref{Ktint}) gives%
\begin{align}
K\left(  t\right)   &  =\int_{-\infty}^{\infty}d\tau K^{WR}\left(
\tau\right)  \left[  \frac{1}{2\pi}\int_{-\infty}^{\infty}d\lambda
e^{-i\lambda\left(  \tau-it\right)  }\right] \nonumber\\
&  =\int_{-\infty}^{\infty}d\tau K^{WR}\left(  \tau\right)  \delta\left(
\tau-it\right)  .
\end{align}
Then, we arrive at a relation between the heat kernel and the Wick-rotated
heat kernel:%
\begin{equation}
K\left(  t\right)  =K^{WR}\left(  it\right)  . \label{Kgenint}%
\end{equation}
This is just the result given by the integral form of the definition of the
generalized heat kernel, Eq. (\ref{genKt}).

In a word, the key idea of introducing the generalized heat kernel for lower
unbounded spectra is an analytic continuation treatment through a Wick rotation.

\subsection{Generalized spectral counting function \label{countingfunction}}

The spectral counting function is the number of eigenstates whose eigenvalue
is smaller than a given number $\lambda$ \cite{dai2009number,dai2010approach},%
\begin{equation}
N\left(  \lambda\right)  =\sum_{\lambda_{0}}^{\lambda_{n}\leq\lambda}1,
\label{cfsum1}%
\end{equation}
where $\lambda_{0}$ denotes the minimum eigenvalue. The spectral counting
function is an important spectral function, which is the starting point of the
famous problem formulated by Kac \textquotedblright Can one hear the shape of
a drum?\textquotedblright\ \cite{kac1966can} The spectral counting function
$N\left(  \lambda\right)  $ has a directly relation with the global heat
kernel $K\left(  t\right)  $, or, the partition function $Z\left(
\beta\right)  $ \cite{dai2009number}. In probability theory, the spectral
counting function corresponds to the cumulative probability function.

For lower unbounded spectra, as that of heat kernels, the definition
(\ref{cfsum1}) needs to be generalized as
\begin{equation}
N\left(  \lambda\right)  =\sum_{-\infty}^{\lambda_{n}\leq\lambda}1.
\label{cfsum}%
\end{equation}
At first sight, this seems to be a wrong definition: if the spectrum is not
lower bounded, usually, there are an infinite number of states below the state
with eigenvalue $\lambda_{n}<\lambda$.

There are two ways to generalize the definition of the spectral counting
function to lower unbounded spectra.

\subsubsection{State density approach}

Directly converting the sum in Eq. (\ref{cfsum}) into an integral is the most
straightforward way to generalize the spectral counting function to lower
unbounded spectra:%
\begin{equation}
N\left(  \lambda\right)  =\int_{-\infty}^{\lambda}\rho\left(  \lambda\right)
d\lambda. \label{NLint}%
\end{equation}
Similarly to heat kernels, the divergence encountered in the sum is avoided by
the treatment of converting the sum into an integral. In this sense, the
counting function is well-defined so long as the state density $\rho\left(
\lambda\right)  $ is integrable.

\subsubsection{Heat kernel approach}

Alternatively, in Refs. \cite{dai2009number,zhou2018calculating}, we present a
relation between heat kernels and counting functions for lower bounded
spectra: $\frac{K\left(  t\right)  }{t}=\int_{0}^{\infty}N\left(
\lambda\right)  e^{-\lambda t}d\lambda$, i.e., the counting function $N\left(
\lambda\right)  $ is a Laplace transformation of $K\left(  t\right)  /t$. For
lower unbounded spectra, the lowest eigenvalue is $-\infty$, so, as that of
heat kernel, we can directly generalize the relation between $K\left(
t\right)  $ and $N\left(  \lambda\right)  $ as%
\begin{equation}
\frac{K\left(  t\right)  }{t}=\int_{-\infty}^{\infty}N\left(  \lambda\right)
e^{-\lambda t}d\lambda. \label{KNtwoL}%
\end{equation}
Clearly, this generalized counting function for lower unbounded spectra is a
two-sided Laplace transformation (bilateral Laplace transformation) of
$K\left(  t\right)  /t$ rather than a Laplace transformation.

Performing a Wick rotation to Eq. (\ref{KNtwoL}), $t=-i\tau$, gives
\begin{equation}
\frac{K\left(  -i\tau\right)  }{-i\tau}=\int_{-\infty}^{\infty}N\left(
\lambda\right)  e^{i\lambda\tau}d\lambda=\mathcal{F}\left[  N\left(
\lambda\right)  ;\tau\right]  .
\end{equation}
Then, the two-sided Laplace transformation in Eq. (\ref{KNtwoL}) is converted
to a Fourier transformation. Therefore, the counting function $N\left(
\lambda\right)  $ can be immediately obtained by an inverse Fourier
transformation:%
\begin{align}
N\left(  \lambda\right)   &  =\mathcal{F}^{-1}\left[  \frac{K\left(
-i\tau\right)  }{-i\tau};\lambda\right] \nonumber\\
&  =\frac{1}{2\pi i}\int_{-\infty}^{\infty}\frac{K\left(  \tau\right)  }{\tau
}e^{i\lambda\tau}d\tau. \label{NKtwoL}%
\end{align}

Eqs. (\ref{KNtwoL}) and (\ref{NKtwoL}) are relations between generalized heat
kernels and generalized counting functions.

It should be noted that, at first sight, the counting function of a lower
unbounded spectrum will diverge since there are infinite eigenvalues below the
given number $\lambda$. However, if instead the naive definition of the
counting function, $N\left(  \lambda\right)  =\sum_{\lambda_{n}<\lambda}1$, by
the definition (\ref{NLint}), we may arrive at a finite result in the case of probability.

\subsection{Generalized spectral zeta function \label{zeta}}

The spectral zeta function is important in quantum field theory, which is the
basics in calculating the one-loop effective action and the vacuum energy,
etc. \cite{vassilevich2003heat,dai2010approach}. In probability, the spectral
zeta function indeed corresponds to the logarithmic moment-generating function.

The conventional definition of the spectral zeta function, which is valid only
for lower bounded spectra, reads%
\begin{equation}
\zeta\left(  s\right)  =\sum_{n=0}^{\infty}\lambda_{n}^{-s}. \label{zetadef}%
\end{equation}

\subsubsection{State density approach}

To generalize the spectral zeta function to lower unbounded spectra, similarly
to the generalized heat kernel, we convert the sum into an integral with the
state density $\rho\left(  \lambda\right)  $:
\begin{equation}
\zeta\left(  s\right)  =\int_{-\infty}^{\infty}d\lambda\rho\left(
\lambda\right)  \lambda^{-s}. \label{zetaint}%
\end{equation}
This definition is valid so long as state density $\rho\left(  \lambda\right)
\lambda^{-s}$ is integrable.

\subsubsection{Heat kernel approach}

The relation between the conventional heat kernel and the conventional
spectral zeta function is a Mellin transformation \cite{vassilevich2003heat},
\begin{align}
\zeta\left(  s\right)   &  =\frac{1}{\Gamma\left(  s\right)  }\int_{0}%
^{\infty}t^{s-1}K\left(  t\right)  dt\nonumber\\
&  =\frac{1}{\Gamma\left(  s\right)  }\mathcal{M}\left[  K\left(  t\right)
;s\right]  , \label{zeta-hk}%
\end{align}
where $\mathcal{M}\left[  K\left(  t\right)  ;s\right]  $ denotes the Mellin
transformation. It can be checked that such a relation also holds for
generalized heat kernels and generalized spectral zeta functions. Substituting
the generalized heat kernel (\ref{Ktint}) into Eq. (\ref{zeta-hk}) gives
\begin{align}
\zeta\left(  s\right)   &  =\frac{1}{\Gamma\left(  s\right)  }\int_{0}%
^{\infty}dtt^{s-1}\left[  \int_{-\infty}^{\infty}d\lambda\rho\left(
\lambda\right)  e^{-\lambda t}\right] \nonumber\\
&  =\int_{-\infty}^{\infty}d\lambda\rho\left(  \lambda\right)  \lambda^{-s}.
\label{spectrumZeta}%
\end{align}

\subsubsection{Characteristic function approach \label{Characteristic}}

First introduce a spectral characteristic function defined as the Fourier
transformation of the state density,
\begin{equation}
f\left(  k\right)  =\int_{-\infty}^{\infty}\rho\left(  \lambda\right)
e^{ik\lambda}d\lambda. \label{characteristic}%
\end{equation}

Constructing a representation of $\lambda^{-s}$,%

\begin{equation}
\lambda^{-s}=\frac{1}{i^{s}\Gamma\left(  s\right)  }\int_{0}^{\infty
}e^{ik\lambda}k^{s-1}dk=\frac{1}{i^{s}\Gamma\left(  s\right)  }\mathcal{M}%
\left[  e^{ik\lambda};s\right]  ,
\end{equation}
and substituting into Eq. (\ref{zetaint}) give%
\begin{align}
\zeta\left(  s\right)   &  =\int_{-\infty}^{\infty}\rho\left(  \lambda\right)
\left[  \frac{1}{i^{s}\Gamma\left(  s\right)  }\int_{0}^{\infty}e^{ik\lambda
}k^{s-1}dk\right]  d\lambda\nonumber\\
&  =\frac{1}{i^{s}\Gamma\left(  s\right)  }\int_{0}^{\infty}\left[
\int_{-\infty}^{\infty}\rho\left(  x\right)  e^{ikx}dx\right]  k^{s-1}dk.
\end{align}
By the definition of the spectral characteristic function
(\ref{characteristic}), we arrive at%
\begin{align}
\zeta\left(  s\right)   &  =\frac{1}{i^{s}\Gamma\left(  s\right)  }\int%
_{0}^{\infty}f\left(  k\right)  k^{s-1}dk\label{zetafk}\\
&  =\frac{1}{i^{s}\Gamma\left(  s\right)  }\mathcal{M}\left[  f\left(
k\right)  ;s\right]  . \label{zetamellin}%
\end{align}

\subsubsection{Mellin transformation approach}

The spectral zeta can also be expressed as a Mellin transformation,%
\begin{equation}
\mathcal{M}\left[  f\left(  t\right)  ;s\right]  =\int_{0}^{\infty}f\left(
t\right)  t^{s-1}dt.
\end{equation}
Rewriting the expression of the spectral zeta function (\ref{zetaint}) as%
\begin{equation}
\zeta\left(  s\right)  =\int_{-\infty}^{0}\rho\left(  x\right)  x^{-s}%
dx+\int_{0}^{\infty}\rho\left(  x\right)  x^{-s}dx,
\end{equation}
we can represent the zeta by the Mellin transformation as%
\begin{equation}
\zeta\left(  s\right)  =\left(  -1\right)  ^{-s-1}\mathcal{M}\left[
-x\rho\left(  -x\right)  ;-s\right]  +\mathcal{M}\left[  x\rho\left(
x\right)  ;-s\right]  .
\end{equation}

\section{Probability spectral function \label{Probabilityspectralfunction}}

The main aim of the present paper is to introduce probability spectral
functions, which bridges spectral theory and probability theory.

The key idea to construct a probability spectral theory, including probability
thermodynamics and probability quantum fields, is to regard random variables
as eigenvalues.

In probability theory, there are two kinds of probability distributions:

(1) the random variable ranges from $0$ to $\infty$;

(2) the random variable ranges from $-\infty$ to $\infty$.\newline When
regarding random variables as eigenvalue spectra, the distribution with random
variables ranging from $0$ to $\infty$ corresponds to the conventional heat
kernel which is appropriate for lower bounded spectra, and the distribution
with random variables ranging from $-\infty$ to $\infty$ corresponds to the
generalized heat kernel which is appropriate for lower unbounded spectra.

After bridging probability theory and spectral theory by regarding random
variables as eigenvalues, we can further\ construct a fictional physical
system whose eigenvalues obey a probability distribution. Thermodynamics and
quantum fields of such a system can be established.

Concretely, for a probability distribution, by regarding the eigenvalue
$\lambda$ as the random variables $x$, regarding the state density
$\rho\left(  \lambda\right)  $ as the probability distribution function
$p\left(  x\right)  $, and regarding the counting function as the cumulative
probability functions $P\left(  x\right)  $, etc., we arrive at a set of
probability spectral functions. In probability theory, there are various
probability distributions, such as the Gaussian distribution, the Laplace
distribution, the student's $t$-distribution, etc. For each probability
distribution, we can construct a family of spectral functions, including,
e.g., the thermodynamic quantity and the effective action.

\subsection{Spectral counting function and cumulative probability function}

In this section, we show that the similarity between spectral counting
functions and\ the cumulative probability functions bridges spectral theory
and probability theory.

The spectral counting function is the number of eigenstates with eigenvalues
less than or equal to $\lambda$ \cite{dai2009number,zhou2018calculating}. In
section \ref{countingfunction}, we generalize the definition of the spectral
counting function to lower-unbounded spectra. The generalized spectral
counting function can be expressed as%
\begin{equation}
N\left(  \lambda\right)  =\int_{\lambda_{\min}}^{\lambda}\rho\left(
\lambda\right)  d\lambda, \label{NLmin}%
\end{equation}
where $\rho\left(  \lambda\right)  $ is the state density and $\lambda_{\min}$
tends to $-\infty$ for lower-unbounded spectra.

In probability theory, there is a cumulative distribution function ------ the
probability that a random variable will be found at a value less than or equal
to $\lambda$ \cite{grimmett2014probability}. The cumulative probability
function can be expressed as%
\begin{equation}
P\left(  x\right)  =\int_{x_{\min}}^{x}p\left(  x\right)  dx, \label{CPF}%
\end{equation}
where $p\left(  x\right)  $ is the probability density function. The random
variable $x_{\min}$ tends to $-\infty$ for probability distributions with
negative infinite random variables.

Comparing the definition of the spectral counting function (\ref{NLmin}) and
the cumulative probability functions (\ref{CPF}), we can see that when
regarding the random variable as an eigenvalue, the cumulative probability
function serves as a spectral counting function.

Along this line of thought, furthermore, we can also find the similarity
between the state density and the probability density function; the
probability density function plays the same role as the state density.

The above observation builds a bridge between the probability theory and the
spectral theory. In the following, by regarding the random variable as an
eigenvalue, we transform the probability theory into a spectral theory.

\subsection{Generalized heat kernel as generalized moment-generating function}

The moment-generating function is an alternative description in addition to
probability density functions and cumulative distribution functions. The
moment-generating function, however, unlike the characteristic function, does
not always exist. In this section, we show that the generalized heat kernel
can serve as a generalized moment-generating function which is definable when
the original definition of the moment-generating function fails.

In probability theory, the moment-generating function $M\left(  t\right)  $ is
defined by a Riemann-Stieltjes integral \cite{grimmett2014probability},%
\begin{equation}
M\left(  t\right)  =\int_{-\infty}^{\infty}e^{tx}dP\left(  x\right)  ,
\end{equation}
where $P\left(  x\right)  $ is the cumulative probability function.

For continuous random variables, the probability density function is $p\left(
x\right)  $. The moment-generating function $M\left(  -t\right)  $ is then a
two-sided Laplace transformation of $p\left(  x\right)  $,
\begin{equation}
M\left(  -t\right)  =\int_{-\infty}^{\infty}p\left(  x\right)  e^{-tx}dx.
\label{ghk}%
\end{equation}

Observing the relation between the generalized heat kernel and the state
density, Eq. (\ref{Ktint}), we can see that the generalized heat kernel
$K\left(  t\right)  =\int_{-\infty}^{\infty}d\lambda\rho\left(  \lambda
\right)  e^{-\lambda t}$ is just the moment-generating function (\ref{ghk}).

When $\rho\left(  \lambda\right)  $ decreases faster than $e^{-\lambda t}$,
the moment-generating function is ill-defined. Nevertheless, even for such a
case, the generalized heat kernel is still well-defined after an analytic
continuation treatment.

In a word, when replacing the moment-generating function with the heat kernel,
we, in fact, introduce a generalized moment-generating function instead of the
original definition.

The moment is fundamentally important in statistics: the first moment is the
mean value, the second central moment is the variance, the third central
moment is the skewness, and the fourth central moment is the kurtosis
\cite{john1988mathematical}. Moreover, the mean value of a quantity can be
expressed as a series of moments by expanding the quantity as a power series.
Nevertheless, many statistical distributions have no moment, such as the
Cauchy distribution and the intermediate distribution
\cite{liu2012intermediate}.

By the generalized moment-generating function, we can define moments for the
statistical distributions which have no moments. In Ref.
\cite{liu2012intermediate}, for defining moments for statistical
distributions, one introduces a weighted moment. When introducing the weighted
moment, the moment depends not only on the distribution but also on the choice
of weighted function. The moment defined in the present paper is more natural,
which depends only on the statistical distribution itself.

\subsection{Generalized heat kernel and characteristic function}

In this section, we present the relation between the generalized heat kernel
in spectral theory and the characteristic function in probability theory.

In probability theory, the characteristic function is defined as the Fourier
transformation of the distribution function $p\left(  x\right)  $
\cite{john1988mathematical}:
\begin{equation}
f\left(  k\right)  =\mathcal{F}\left[  p\left(  x\right)  ;k\right]
=\int_{-\infty}^{\infty}dxp\left(  x\right)  e^{ikx}, \label{fkdef}%
\end{equation}
where $\mathcal{F}\left[  p\left(  x\right)  ;k\right]  $ denotes the Fourier
transformation of $p\left(  x\right)  $.

The relation between the generalized heat kernel and the characteristic
function can be obtained by inspection of their definitions (\ref{Ktint}) and
(\ref{fkdef}).

As discussed above, by regarding the statistical distribution function
$p\left(  x\right)  $ as a state density $\rho\left(  \lambda\right)  $ of a
spectrum, we have%
\begin{equation}
K\left(  t\right)  =\int_{-\infty}^{\infty}dxp\left(  x\right)  e^{-xt}.
\label{Ktpx}%
\end{equation}
From the definition of the characteristic function (\ref{fkdef}), we can see
that the statistical distribution function $p\left(  x\right)  $ is the
inverse Fourier transformation of the characteristic function:%
\begin{equation}
p\left(  x\right)  =\frac{1}{2\pi}\int_{-\infty}^{\infty}dkf\left(  k\right)
e^{-ikx}. \label{pxfk}%
\end{equation}
Substituting Eq. (\ref{pxfk}) into Eq. (\ref{Ktpx}) gives%
\begin{align}
K\left(  t\right)   &  =\int_{-\infty}^{\infty}dkf\left(  k\right)  \frac
{1}{2\pi}\int_{-\infty}^{\infty}dxe^{-i\left(  k-it\right)  x}\nonumber\\
&  =\int_{-\infty}^{\infty}dkf\left(  k\right)  \delta\left(  k-it\right)  .
\end{align}
Then, we arrive at a relation between the generalized heat kernel and the
characteristic function:%
\begin{equation}
K\left(  t\right)  =f\left(  it\right)  . \label{HKC}%
\end{equation}
It can be directly seen from Eq. (\ref{Kgenint}) that the characteristic
function is just the Wick rotated heat kernel introduced in section
\ref{intdefintion}.

\subsection{Probability spectral zeta function}

Similarly, we can construct the probability zeta function. From the
generalized spectral zeta function (\ref{spectrumZeta}),\ by regarding the
eigenvalue $\lambda$ as the random variables $x$ and regarding the statistical
distribution function $p\left(  x\right)  $ as the state density $\rho\left(
\lambda\right)  $, we arrive at a probability zeta function.
\begin{equation}
\zeta\left(  s\right)  =\int_{-\infty}^{\infty}dxp\left(  x\right)  x^{-s}.
\label{zetaP}%
\end{equation}

Based on this probability zeta function, we can obtain various spectral
functions in quantum field theory, e.g., the effective action.

\section{Probability thermodynamics \label{Thermodynamics}}

All thermodynamic quantities are spectral functions. The thermodynamic
behavior is determined by the spectrum. Starting from the partition function
$Z\left(  \beta\right)  $, whether the conventional definition or the
generalized definition, we arrive at thermodynamics. All thermodynamic
quantities can be obtained from the partition function directly.

Regarding the probability density function as a state density of an eigenvalue
spectrum, each probability distribution, such as the Gaussian distribution and
the Cauchy distribution, defines a thermodynamic system. This allows us to
establish various probability thermodynamics corresponding to various
probability distributions.

\subsection{Probability partition function}

In statistical mechanics, the canonical partition function is defined by%
\begin{equation}
Z\left(  \beta\right)  =\sum_{\left\{  \lambda_{n}\right\}  }e^{-\beta
\lambda_{n}}. \label{Zdfn}%
\end{equation}
The canonical partition function is just the heat kernel with the replacement
of $t$ by $\beta$.

For the same reason as in the definition of the heat kernel, the definition of
the canonical partition function is only valid for lower bounded spectra. For
lower unbounded spectra, as done for heat kernels, the canonical partition
function can also be generalized.

According to the generalization of the heat kernel given above, by Eqs.
(\ref{Kgenint}) and (\ref{KtWR}), we immediately achieve a generalized
partition function,
\begin{equation}
Z\left(  \beta\right)  =\left.  \int_{\lambda_{\min}}^{\infty}\rho\left(
\lambda\right)  e^{-i\lambda\tau}d\lambda\right\vert _{\tau=-i\beta};
\label{ZbetaWR}%
\end{equation}
or, by Eq. (\ref{Ktint}), we arrive at%

\begin{equation}
Z\left(  \beta\right)  =\int_{\lambda_{\min}}^{\infty}d\lambda\rho\left(
\lambda\right)  e^{-\beta\lambda}. \label{Zbeta}%
\end{equation}
For lower unbounded spectra, $\lambda_{\min}\rightarrow-\infty$.

It can be directly seen from Eq. (\ref{Zbeta}) that, if regarding $\rho\left(
\lambda\right)  e^{-\beta\lambda}$ as a density function, the partition
function is the zero-order moment $\left\langle \lambda^{0}\right\rangle $.

It should be emphasized that the partition function given by Eq.
(\ref{ZbetaWR}) or (\ref{Zbeta}) is a generalized partition function rather
than the conventional partition function, and the corresponding thermodynamic
quantities are also not the conventional thermodynamic quantities.

\subsection{Probability thermodynamic quantity}

Starting from the generalized canonical partition function, Eq. (\ref{ZbetaWR}%
) or (\ref{Zbeta}), we can construct whole thermodynamics regardless of the
spectrum lower bounded or lower unbounded.

The internal energy is a statistical average of eigenvalues formally defined
by
\begin{equation}
U\left(  \beta\right)  =\frac{\sum_{\left\{  \lambda_{n}\right\}  }\lambda
_{n}e^{-\lambda_{n}\beta}}{\sum_{\left\{  \lambda_{n}\right\}  }%
e^{-\lambda_{n}\beta}}. \label{Udfn}%
\end{equation}

For lower unbounded spectra, such a definition becomes $U\left(  \beta\right)
=\frac{\sum_{-\infty}^{\infty}\lambda_{n}e^{-\lambda_{n}\beta}}{\sum_{-\infty
}^{\infty}e^{-\lambda_{n}\beta}}$. The divergence here is no longer a problem,
for it can be removed by the method discussed above. It is worth pointing out
that the internal energy for lower unbounded spectra may be negative since
part of the eigenvalues can take negative values.

Observing the definition of the internal energy (\ref{Udfn}) and\ the
partition function (\ref{Zdfn}), we can formally write down the relation of
internal energy and partition function even for lower unbounded spectra
\begin{equation}
U\left(  \beta\right)  =-\frac{\partial}{\partial\beta}\ln Z\left(
\beta\right)  =\frac{1}{Z\left(  \beta\right)  }\int_{\lambda_{\min}}^{\infty
}d\lambda\rho\left(  \lambda\right)  \lambda e^{-\beta\lambda}. \label{Udef}%
\end{equation}

From Eq. (\ref{Udef}), we can see that the internal energy is the first-order
moment $\left\langle \lambda^{1}\right\rangle $.

Furthermore, the specific heat can be calculated from the internal energy:%
\begin{equation}
C_{V}=\frac{\partial U}{\partial T}=\beta^{2}\left[  \frac{\int_{\lambda
_{\min}}^{\infty}d\lambda\rho\left(  \lambda\right)  \lambda^{2}%
e^{-\lambda\beta}}{\int_{\lambda_{\min}}^{\infty}d\lambda\rho\left(
\lambda\right)  e^{-\lambda\beta}}-\left(  \frac{\int_{\lambda_{\min}}%
^{\infty}d\lambda\rho\left(  \lambda\right)  \lambda e^{-\lambda\beta}}%
{\int_{\lambda_{\min}}^{\infty}d\lambda\rho\left(  \lambda\right)
e^{-\lambda\beta}}\right)  ^{2}\right]  .
\end{equation}
It can be seen that the specific heat is the difference between the
second-order moment and the square of the first-order moment $\left\langle
\lambda^{2}\right\rangle -\left\langle \lambda^{1}\right\rangle ^{2}$.

Moreover, from the generalized canonical partition function, we can obtain
other thermodynamic quantities, e.g., the free energy%
\begin{equation}
F\left(  \beta\right)  =-\frac{1}{\beta}\ln Z\left(  \beta\right)
\end{equation}
and the entropy%
\begin{equation}
S=\ln Z\left(  \beta\right)  -\beta\frac{\partial}{\partial\beta}\ln Z\left(
\beta\right)
\end{equation}
for both lower bounded and lower unbounded spectra.

\subsection{Probability thermodynamics: characteristic function approach}

Starting from the relation between the heat kernel and the characteristic
function, we can directly represent various thermodynamic quantities by the
characteristic function. According to the correspondence between the
generalized heat kernel and the canonical partition function (\ref{Zbeta}) and
the relation (\ref{HKC}), we represent the canonical partition function as
\begin{equation}
Z\left(  \beta\right)  =f\left(  i\beta\right)  .
\end{equation}
Various thermodynamic quantities then can also be represented by the
characteristic function, e.g., the internal energy%
\begin{equation}
U\left(  \beta\right)  =-\frac{\partial}{\partial\beta}\ln f\left(
i\beta\right)
\end{equation}
and the specific heat capacity%
\begin{equation}
C_{V}=\beta^{2}\frac{\partial^{2}}{\partial\beta^{2}}\ln f\left(
i\beta\right)  ,
\end{equation}
etc.

\section{Probability quantum field theory \label{QFT}}

The vacuum amplitude in the Euclidean quantum field theory is%
\begin{equation}
Z=\int\mathcal{D}\phi e^{-I\left[  \phi\right]  /\hbar},
\label{Cvacuumapmlitude}%
\end{equation}
where $I\left[  \phi\right]  $ is the Euclidean action $I\left[  \phi\right]
=-\int d^{3}xd\left(  it\right)  \mathcal{L}$. The Euclidean vacuum amplitude
in spectral representation is just the partition function or the global heat
kernel (\ref{trdht}).

The conventional definition of the vacuum amplitude (\ref{Cvacuumapmlitude})
is, of course, only valid for lower bounded spectra. This procedure can also
provide a generalized definition of vacuum amplitude for lower unbounded
spectra. This allows us to construct a quantum field for lower unbounded operators.

By regarding the probability partition function as a probability vacuum
amplitude, we arrive at a probability quantum field. Taking the one-loop
effective action and the vacuum energy as examples, we show how to construct a
probability quantum field.

\subsection{One-loop effective action and vacuum energy as spectral function}

The information of a mechanical system is embedded in an Hermitian operator
$D$. Many important physical quantities are spectral functions constructed
from the eigenvalues $\left\{  \lambda_{n}\right\}  $ of the operator $D$. Two
important quantities in quantum field theory, the one-loop effective action
and the vacuum energy, are both spectral functions: the one-loop effective
action is the determinant of the operator, $\det D$, and the vacuum energy is
the trace of the operator, $\operatorname{tr}D$.

For an Hermitian operator, the determinant is the product of the eigenvalues:
\begin{equation}
\det D=%
%TCIMACRO{\dprod _{n}}%
%BeginExpansion
{\displaystyle\prod_{n}}
%EndExpansion
\lambda_{n},
\end{equation}
and the trace is the sum of the eigenvalues:
\begin{equation}
\operatorname{tr}D=%
%TCIMACRO{\dsum _{n}}%
%BeginExpansion
{\displaystyle\sum_{n}}
%EndExpansion
\lambda_{n}. \label{trD}%
\end{equation}
They are, obviously, divergent.

In order to obtain a finite result, we need a renormalization treatment.

\subsection{Probability one-loop effective action}

The effective action for operator $D$ can be expanded as%
\begin{equation}
\Gamma\left[  \phi\right]  =I\left[  \phi\right]  +\hbar\frac{1}{2}\ln\det
D+O\left(  \hbar^{2}\right)  ,
\end{equation}
where
\begin{equation}
W=\frac{1}{2}\ln\det D=\frac{1}{2}\ln%
%TCIMACRO{\dprod \limits_{n}}%
%BeginExpansion
{\displaystyle\prod\limits_{n}}
%EndExpansion
\lambda_{n}=\frac{1}{2}\sum_{n}\ln\lambda_{n} \label{Woneloop}%
\end{equation}
is the one-loop effective action which diverges.

In order to obtain a finite one-loop effective action, we analytically
continue the sum in the one-loop effective action (\ref{Woneloop}) by virtue
of the spectral zeta function. By Eq. (\ref{zetadef}) we can see that the
derivative of the spectral zeta function is $\zeta^{\prime}\left(  s\right)
=-\sum_{n}\lambda_{n}^{-s}\ln\lambda_{n}$, so $\zeta^{\prime}\left(  0\right)
=\sum_{n}\ln\lambda_{n}$ and then
\begin{equation}
W=-\frac{1}{2}\zeta^{\prime}\left(  0\right)  . \label{Wac}%
\end{equation}

In practice, the one-loop effective action for continuous spectra can be
rewritten as%
\begin{equation}
W=-\frac{1}{2}\int_{-\infty}^{\infty}\rho\left(  \lambda\right)  \ln\lambda
d\lambda, \label{Wint}%
\end{equation}
by use of $\zeta^{\prime}\left(  0\right)  =\sum_{n}\ln\lambda_{n}%
=\int_{-\infty}^{\infty}\rho\left(  \lambda\right)  \ln\lambda d\lambda$.

Now, we show that, in probability quantum field theory, the expression
(\ref{Wac}) for one-loop effective action is often convergent, though it
diverges in many cases of quantum field theory.

In quantum field theory, in order to remove the divergence, one introduces a
regularized one-loop effective action \cite{vassilevich2003heat}:%
\begin{equation}
W_{s}=-\frac{1}{2}\widetilde{\mu}^{2s}\Gamma\left(  s\right)  \zeta\left(
s\right)  .
\end{equation}
where $\widetilde{\mu}$ is a constant of the dimension of mass introduced to
keep the proper dimension of the effective action. To remove the divergence in
the one-loop effective action, we expand $W_{s}$ around $s=0$:%
\begin{equation}
W_{s}=-\frac{1}{2}\zeta\left(  0\right)  \frac{1}{s}-\frac{1}{2}\zeta^{\prime
}\left(  0\right)  +\left(  \frac{1}{2}\gamma_{E}-\ln\widetilde{\mu}\right)
\zeta\left(  0\right)  .
\end{equation}
After the minimal subtraction, which drops the divergent term, we arrive at a
renormalized one-loop effective action,%
\begin{equation}
W^{\text{ren}}=-\frac{1}{2}\zeta^{\prime}\left(  0\right)  +\left(  \frac
{1}{2}\gamma_{E}-\ln\widetilde{\mu}\right)  \zeta\left(  0\right)  .
\end{equation}
By the definition (\ref{zetadef}), $\zeta\left(  0\right)  =\sum_{n}1$ is
divergent. It is just the divergent $\zeta\left(  0\right)  $ cancels the
divergence in $\zeta^{\prime}\left(  0\right)  $.

In probability quantum field theory, the spectra satisfy probability
distributions, so $\zeta\left(  0\right)  =\sum_{n}1=\int_{-\infty}^{\infty
}\rho\left(  \lambda\right)  d\lambda=1$ is the total probability which is of
course finite and normalized. For finite $\zeta\left(  0\right)  $, choosing
the constant $\widetilde{\mu}^{2}=e^{\gamma_{E}}$, we return to Eq. (\ref{Wac}).

Moreover, it should be emphasized that in probability quantum field theory,
$\zeta^{\prime}\left(  0\right)  $\ still may diverge in some cases. For
continuous spectra, $\zeta^{\prime}\left(  0\right)  =\sum_{n}\ln\lambda
_{n}=\int_{-\infty}^{\infty}\rho\left(  \lambda\right)  \ln\lambda d\lambda$.
In probability theory, $\rho\left(  \lambda\right)  $\ is the probability
distribution function. Obviously, for probability distribution possessing
moments, $\zeta^{\prime}\left(  0\right)  $\ is finite. For probability
distribution without moments, we need to analyze the integrability case by case.

The one-loop effective action can also be calculated directly from the
spectral characteristic function introduced in section \ref{Characteristic}.

Representing the state density $\rho\left(  \lambda\right)  $ by the inverse
Fourier transformation of the spectral characteristic function, by Eq.
(\ref{pxfk}),%
\begin{equation}
\rho\left(  \lambda\right)  =\frac{1}{2\pi}\int_{-\infty}^{\infty}f\left(
k\right)  e^{ik\lambda}dk
\end{equation}
and substituting into the expression of the one-loop effective action, Eq.
(\ref{Wint}), we arrive at%
\begin{align}
W  &  =\int_{-\infty}^{\infty}\left[  \frac{1}{2\pi}\int_{-\infty}^{\infty
}f\left(  k\right)  e^{ik\lambda}dk\right]  \ln\lambda d\lambda\nonumber\\
&  =\frac{1}{2\pi}\int_{-\infty}^{\infty}f\left(  k\right)  \left[
\int_{-\infty}^{\infty}e^{ik\lambda}\ln\lambda d\lambda\right]  dk.
\label{Wfk}%
\end{align}
The integral in Eq. (\ref{Wfk}) diverges. This integral, however, is a Fourier
transformation of $\ln\lambda$. After analytic continuation, we achieve a
finite result:
\begin{equation}
\int_{-\infty}^{\infty}e^{ik\lambda}\ln\lambda d\lambda=\pi\left[
-2\gamma_{\text{E}}\delta\left(  k\right)  +i\pi\delta\left(  k\right)
-\frac{\operatorname*{sgn}\left(  k\right)  -1}{k}\right]  .
\end{equation}
Then we have
\begin{equation}
W=\left(  -\gamma_{\text{E}}+\frac{i\pi}{2}\right)  f\left(  0\right)
-\frac{1}{2}\int_{-\infty}^{\infty}f\left(  k\right)  \frac
{\operatorname*{sgn}\left(  k\right)  -1}{k}dk.
\end{equation}
Therefore,
\begin{equation}
W=\left(  -\gamma_{\text{E}}+\frac{i\pi}{2}\right)  f\left(  0\right)
+\int_{-\infty}^{0}\frac{f\left(  k\right)  }{k}dk. \label{Wfk2}%
\end{equation}

This expression is useful when only the characteristic function of a
probability distribution is known.

In some cases, the integral term in Eq. (\ref{Wfk2}) still diverges. However,
the divergence can be removed by a renormalization procedure
\cite{li2016scattering}. We will illustrate such a renormalization procedure
by taking the Cauchy distribution as an example in section \ref{Cauchy}.

\subsection{Probability vacuum energy \label{E0}}

The vacuum energy is essentially a sum of all eigenvalues or the trace of the
operator $D$:
\begin{equation}
E_{0}=%
%TCIMACRO{\dsum _{n}}%
%BeginExpansion
{\displaystyle\sum_{n}}
%EndExpansion
\lambda_{n}.
\end{equation}
Physical operators are not upper-bounded, so the vacuum energy diverges. In
section \ref{zeta}, we suggest an approach to obtain a finite result of the
trace of the operator.

The\ trace of the operator $D$, Eq. (\ref{trD}), is a Dirichlet series defined
by $%
%TCIMACRO{\dsum _{n=1}^{\infty}}%
%BeginExpansion
{\displaystyle\sum_{n=1}^{\infty}}
%EndExpansion
\frac{a_{n}}{\lambda_{n}^{s}}$ which converges absolutely on the right open
half-plane such that $\operatorname{Re}s>1$. A Dirichlet series can be
analytically continued as a spectral zeta function.

In our problem, the spectrum may not be lower-bounded. That is, the trace of
the operator encountered sometimes is a generalized Dirichlet series. By the
procedure developed in section \ref{zeta}, we can analytically continue such a
generalized Dirichlet series as a generalized spectral zeta function.

Concretely, the vacuum energy is just a spectral zeta function:%

\begin{equation}
E_{0}=\operatorname{tr}D=\left.
%TCIMACRO{\dsum _{n}}%
%BeginExpansion
{\displaystyle\sum_{n}}
%EndExpansion
\frac{1}{\lambda_{n}^{s}}\right\vert _{s=-1}=\left.  \zeta\left(  s\right)
\right\vert _{s=-1}=\zeta\left(  -1\right)  . \label{E0R}%
\end{equation}
The trace, after being analytically continued as a spectral zeta function,
$\zeta_{D}\left(  -1\right)  $, is already finite. That is, it is renormalized.

Similarly, in quantum field theory, the vacuum energy always diverges.
However, it recovers $\zeta\left(  -1\right)  $ after a renormalization treatment.

The vacuum energy can also be calculated from the characteristic function.

The vacuum energy can be expressed by the state density,
\begin{equation}
E_{0}=\int_{-\infty}^{\infty}\rho\left(  \lambda\right)  \lambda d\lambda.
\end{equation}
The state density can be expressed by the inverse Fourier transformation of
the characteristic function, so by Eq. (\ref{pxfk}),%
\begin{align}
E_{0}  &  =\left(  -i\right)  \left.  \frac{d}{dk}\left[  \int_{-\infty
}^{\infty}\rho\left(  \lambda\right)  e^{ik\lambda}d\lambda\right]
\right\vert _{k=0}\nonumber\\
&  =\left(  -i\right)  \left.  \frac{df\left(  k\right)  }{dk}\right\vert
_{k=0}. \label{E0ch}%
\end{align}

\section{Probability thermodynamics and probability quantum field: examples
\label{various}}

Following the approach introduced above, we can construct probability
thermodynamics and probability quantum fields for various statistical distributions.

There are two kinds of probability distributions: lower bounded and lower unbounded.

For distributions with nonnegative random variables, such as the gamma
distribution, the exponential distribution, the beta distribution, the
chi-square distribution, the Pareto distribution, the generalized inverse
Gaussian distribution, the tempered stable distribution,\ and the log-normal
distribution, we can use the conventional definition of the spectral function.

For distributions with negative random variables, such as the normal
distribution, the skew Gaussian distribution, Student's $t$-distribution, the
Laplace distribution, the Cauchy distribution, the $q$-Gaussian distribution,
the intermediate distribution, we must use the generalized definition of the
spectral function.

\subsection{Distribution with nonnegative random variable: Gamma distribution}

For distributions with nonnegative random variables, probability
thermodynamics and probability quantum fields can be constructed by the
conventional definition of spectral functions.

In this section, we use the Gamma distribution to illustrate the construction
of thermodynamics and quantum field theory for the distribution with
nonnegative random variables. The thermodynamics and quantum field theory of
other distributions with nonnegative random variables are presented in
Appendix \ref{nonnegative}.

The Gamma distribution is used to describe nonstationary earthquake activity,
the drop size in sprays \cite{michas2018stochastic}, and maximum entropy
approach to H-theory \cite{vasconcelos2018maximum}, etc.

The gamma distribution \cite{ross2014introduction}%
\begin{equation}
p\left(  x\right)  =\frac{1}{\Gamma\left(  a\right)  }b^{-a}e^{-x/b}x^{a-1},
\end{equation}
\ recovers many distributions when taking the shape parameter $a$ and the
inverse scale parameter $b$ as some special values. For example, the
exponential distribution is recovered when $a=1$ and $b=1/\lambda$\ and the
normal distribution is recovered when $a\rightarrow\infty$.

The cumulative distribution function, by Eq. (\ref{CPF}), reads%
\begin{equation}
P\left(  x\right)  =Q\left(  a,0,\frac{x}{b}\right)  ,
\end{equation}
where $Q\left(  a,z_{0},z_{1}\right)  =\Gamma\left(  a,z_{0},z_{1}\right)
/\Gamma\left(  a\right)  $ is the generalized regularized incomplete gamma
function with $\Gamma\left(  a,z_{0},z_{1}\right)  $ the generalized
incomplete gamma function \cite{abramowitz1964handbook}.

\subsubsection{Thermodynamics}

In order to obtain the thermodynamic quantity, we first calculate the
characteristic function by Eq. (\ref{fkdef}), which is a Fourier
transformation of the density function $p\left(  x\right)  $,%
\begin{equation}
f\left(  k\right)  =(1-ibk)^{-a}.
\end{equation}

The heat kernel, by performing the Wick rotation $k\rightarrow it$ to the
characteristic function, reads%
\begin{equation}
K\left(  t\right)  =(1+bt)^{-a}.
\end{equation}

The thermodynamics corresponding to the gamma distribution can be then
constructed. The partition function, by replacing $t$\ with $\beta$\ in the
heat kernel, reads
\begin{equation}
Z\left(  \beta\right)  =(1+b\beta)^{-a}.
\end{equation}
Then the free energy%
\begin{equation}
F=-\frac{1}{\beta}\ln(1+b\beta)^{-a},
\end{equation}
the internal energy%
\begin{equation}
U=\frac{ba}{1+b\beta},
\end{equation}
the entropy%
\begin{equation}
S=\frac{ba\beta}{1+b\beta}-a\ln(1+b\beta),
\end{equation}
and the specific heat%
\begin{equation}
C_{V}=\frac{b^{2}\text{$\beta$}^{2}a}{(1+b\beta)^{2}}.
\end{equation}

\subsubsection{Quantum field}

The spectral zeta function, by Eq. (\ref{zetaP}), reads%
\begin{equation}
\zeta\left(  s\right)  =\frac{1}{\Gamma(a)}b^{-s}\Gamma(a-s).
\end{equation}

The one-loop effective action by Eq. (\ref{Wac}) reads
\begin{equation}
W=\frac{1}{2}\left(  \psi^{(0)}\left(  a\right)  +\ln b\right)  ,
\end{equation}
where $\psi^{(n)}\left(  z\right)  $ is the $n$-th derivative of the digamma function.

The vacuum energy can be obtained by Eq. (\ref{E0R}):%
\begin{equation}
E_{0}=ab.
\end{equation}

\subsection{Distribution with negative random variable: Normal and Cauchy distribution}

In the above, we construct probability thermodynamics and probability quantum
fields for probability distributions with nonnegative random variables. For
constructing probability spectral functions with nonnagative random variables,
we can use the conventional definition of spectral functions. For the
distribution with negative random variables, however, the conventional
definition is invalid, and we need to use the generalized definition given in
the present paper. In this section, we use the Normal distribution and the
Cauchy distribution as examples. The thermodynamics and quantum field theory
of other distributions with negative random variables are presented in
Appendix \ref{negative}.

\subsubsection{Normal distribution}

The normal distribution is also known as the bell curve or the Gaussian distribution.

The probability density function of the normal distribution is
\cite{johnson1994continuous,ross2014introduction}%

\begin{equation}
p\left(  x\right)  =\frac{1}{\sqrt{2\pi}\sigma}\exp\left(  -\frac{(x-\mu)^{2}%
}{2\sigma^{2}}\right)  ,
\end{equation}
where $%
%TCIMACRO{\U{3bc} }%
%BeginExpansion
\mu
%EndExpansion
$ is the mean of the distribution and $\sigma$ is the standard deviation.

The cumulative distribution function, by Eq. (\ref{CPF}), reads%
\begin{equation}
P\left(  x\right)  =\frac{1}{2}\operatorname{erfc}\left(  \frac{\mu-x}%
{\sqrt{2}\sigma}\right)  .
\end{equation}

\paragraph{Thermodynamics}

In order to obtain the thermodynamic quantity, we first calculate the
characteristic function by Eq. (\ref{fkdef}), which is a Fourier
transformation of the density function $p\left(  x\right)  $,%
\[
f\left(  k\right)  =\exp\left(  ik\mu-\frac{k^{2}\sigma^{2}}{2}\right)  .
\]

The heat kernel, by performing the Wick rotation $k\rightarrow it$ to the
characteristic function, reads%
\begin{equation}
K\left(  t\right)  =\exp\left(  \frac{1}{2}t\left(  t\sigma^{2}-2\mu\right)
\right)  .
\end{equation}

The thermodynamics corresponding to the normal distribution can be then
constructed. The partition function, by replacing $t$\ with $\beta$\ in the
heat kernel, reads
\begin{equation}
Z\left(  \beta\right)  =\exp\left(  \frac{1}{2}\beta\left(  \beta\sigma
^{2}-2\mu\right)  \right)  .
\end{equation}
Then the free energy%
\begin{equation}
F=\mu-\frac{1}{2}\beta\sigma^{2},
\end{equation}
the internal energy%
\begin{equation}
U=\mu-\beta\sigma^{2},
\end{equation}
the entropy%
\begin{equation}
S=-\frac{1}{2}\beta^{2}\sigma^{2},
\end{equation}
and the specific heat%
\begin{equation}
C_{V}=\beta^{2}\sigma^{2}.
\end{equation}

\paragraph{Quantum field}

The spectral zeta function, by Eq. (\ref{zetaP}), reads%
\begin{align}
\zeta\left(  s\right)   &  =\frac{\left(  -1\right)  ^{s}}{2^{s/2+1}\sqrt{\pi
}\sigma^{s}}\left\{  \left[  1+(-1)^{s}\right]  \Gamma\left(  \frac{1}%
{2}-\frac{s}{2}\right)  \,_{1}F_{1}\left(  \frac{s}{2};\frac{1}{2};-\frac
{\mu^{2}}{2\sigma^{2}}\right)  \right. \nonumber\\
&  \left.  +\frac{\sqrt{2}\mu\left[  (-1)^{s}-1\right]  }{\sigma}\Gamma\left(
1-\frac{s}{2}\right)  \,_{1}F_{1}\left(  \frac{s+1}{2};\frac{3}{2};-\frac
{\mu^{2}}{2\sigma^{2}}\right)  \right\}  ,
\end{align}
where $\Gamma\left(  z\right)  $ is the Euler gamma function and $_{1}%
F_{1}(\alpha;b;z)$ is the confluent hypergeometric function. When $s$ is an
integer,%
\begin{align}
\zeta\left(  s\right)   &  =\frac{1}{2^{s/2+1}\sqrt{\pi}\sigma^{s}}\left\{
\left[  1+\left(  -1\right)  ^{s}\right]  \Gamma\left(  \frac{1-s}{2}\right)
\,_{1}F_{1}\left(  \frac{s}{2};\frac{1}{2};-\frac{\mu^{2}}{2\sigma^{2}%
}\right)  \right. \nonumber\\
&  \left.  +\frac{\sqrt{2}\mu\left(  -1\right)  ^{s}\left[  1-\left(
-1\right)  ^{s}\right]  }{\sigma}\Gamma\left(  1-\frac{s}{2}\right)
\,_{1}F_{1}\left(  \frac{s+1}{2};\frac{3}{2};-\frac{\mu^{2}}{2\sigma^{2}%
}\right)  \right\}  .
\end{align}
When $s$ is an even number
\begin{equation}
\zeta\left(  s\right)  =\frac{1}{2^{s/2}\sqrt{\pi}\sigma^{s}}\Gamma\left(
\frac{1-s}{2}\right)  \,_{1}F_{1}\left(  \frac{s}{2};\frac{1}{2};-\frac
{\mu^{2}}{2\sigma^{2}}\right)
\end{equation}
and when $s$ is an odd number%
\begin{equation}
\zeta\left(  s\right)  =-\frac{2^{\left(  1-s\right)  /2}\mu}{\sqrt{\pi}%
\sigma^{s+1}}\Gamma\left(  1-\frac{s}{2}\right)  \,_{1}F_{1}\left(  \frac
{s+1}{2};\frac{3}{2};-\frac{\mu^{2}}{2\sigma^{2}}\right)  .
\end{equation}

The one-loop effective action by Eq. (\ref{Wac}) reads
\begin{equation}
W=\frac{1}{4}\left(  \left.  \frac{\partial\text{ }_{1}F_{1}\left(
a,1/2,-\mu^{2}/2\sigma^{2}\right)  }{\partial z}\right\vert _{a=0}%
+i\pi\operatorname{erfc}\left(  \frac{\mu}{\sqrt{2}\sigma}\right)  +\ln
\frac{\sigma^{2}}{2}-\gamma_{\text{E}}\right)  .
\end{equation}
The vacuum energy can be obtained by Eq. (\ref{E0R}):%
\begin{equation}
E_{0}=\mu.
\end{equation}

\subsubsection{Cauchy distribution \label{Cauchy}}

The Cauchy distribution \cite{johnson1994continuous}%
\begin{equation}
p\left(  x\right)  =\frac{\sigma}{\pi\left[  (x-\mu)^{2}+\sigma^{2}\right]  }.
\label{Cauchydistribution}%
\end{equation}
where $\mu$ is location parameter and $\sigma$ is scale parameter.

The cumulative distribution function%
\begin{equation}
P\left(  x\right)  =\frac{1}{\pi}\tan^{-1}\left(  \frac{x-\mu}{\sigma}\right)
+\frac{1}{2}.
\end{equation}

We first calculate thermodynamics and quantum-field quantities by the similar
direct calculation. Then we calculation the one-loop effective by
renormalization approach as a comparison.

\paragraph{Thermodynamics}

The characteristic function%
\begin{equation}
f\left(  k\right)  =e^{ik\mu-\sigma\left\vert k\right\vert }.
\end{equation}

The heat kernel%
\begin{equation}
K\left(  t\right)  =e^{-t\mu-\sigma\left\vert t\right\vert }.
\end{equation}

The partition function
\begin{equation}
Z\left(  \beta\right)  =e^{-\beta(\mu+\sigma)}.
\end{equation}

The free energy%
\begin{equation}
F=\mu+\sigma.
\end{equation}

The internal energy%
\begin{equation}
U=\mu+\sigma.
\end{equation}

\paragraph{Quantum field}

The spectral zeta function%
\begin{equation}
\zeta\left(  s\right)  =\left(  \mu+i\sigma\right)  ^{-s}.
\end{equation}

The one-loop effective action
\begin{equation}
W=\frac{1}{2}\ln(\mu+i\sigma).
\end{equation}

The vacuum energy%
\begin{equation}
E_{0}=\mu+i\sigma.
\end{equation}

\paragraph{Renormalization}

In this section, we show that the renormalization procedure can remove the
divergence appears in the one-loop effective action and the result coincides
with the renormalized one-loop effective action obtain by the standard quantum
field method. A systematic discussion on removing the divergence in divergent
moments in probability theory can be found in Ref.
\cite{zhang2022renormalization}.

Consider a standard Cauchy distribution which is the distribution
(\ref{Cauchydistribution}) with $\mu=0$ and $\sigma=1$,%
\begin{equation}
p\left(  x\right)  =\frac{1}{\pi\left(  x^{2}+1\right)  }. \label{sCauchy}%
\end{equation}
By Eq. (\ref{Wint}), directly calculation gives%
\begin{align}
W  &  =\int_{-\infty}^{\infty}\frac{1}{\pi\left(  \lambda^{2}+1\right)  }%
\ln\lambda d\lambda\nonumber\\
&  =i\frac{\pi}{2}. \label{WCauchy}%
\end{align}

The one-loop effective action can also be obtained by the characteristic
function. The result can be obtained by Eq. (\ref{Wfk2}) with a
renormalization treatment.

Substituting the distribution (\ref{sCauchy}) into Eq. (\ref{Wfk2}) gives
\begin{equation}
W=\left(  -\gamma_{\text{E}}+\frac{i\pi}{2}\right)  f\left(  0\right)
+\int_{-\infty}^{0}\frac{f\left(  k\right)  }{k}dk.
\end{equation}
The characteristic function of the standard Cauchy distribution is%
\begin{equation}
f\left(  k\right)  =e^{-\left\vert k\right\vert },
\end{equation}
so
\begin{equation}
W=-\gamma_{\text{E}}+\frac{i\pi}{2}+\int_{-\infty}^{0}\frac{e^{k}}{k}dk.
\end{equation}
The integral diverges since $k=0$ is singular. Cutting off the upper limit of
the integral gives
\begin{equation}
\int_{-\infty}^{\varepsilon}\frac{e^{k}}{k}dk=\operatorname{Ei}\left(
\varepsilon\right)  ,
\end{equation}
where $\operatorname{Ei}\left(  z\right)  $ is the exponential integral
function. Expanding at $\varepsilon\rightarrow0$, we arrive at
\begin{equation}
\int_{-\infty}^{\varepsilon}\frac{e^{k}}{k}dk=\operatorname{Ei}\left(
\varepsilon\right)  =\gamma_{\text{E}}+\ln\left(  -\varepsilon\right)  .
\end{equation}
Dropping the divergent term $\ln\left(  -\varepsilon\right)  $, we have
\begin{equation}
\left.  \int_{-\infty}^{0}\frac{e^{k}}{k}dk\right\vert _{\text{reg}}%
=\gamma_{\text{E}}.
\end{equation}
Then we obtain a renormalized one-loop effective action%
\begin{align}
\left.  W\right\vert _{\text{reg}}  &  =-\gamma_{\text{E}}+\frac{i\pi}%
{2}+\left.  \int_{-\infty}^{0}\frac{e^{k}}{k}dk\right\vert _{\text{reg}%
}\nonumber\\
&  =\frac{i\pi}{2},
\end{align}
which agrees with Eq. (\ref{WCauchy}).

The validity of the above renormalization procedure is discussed in Ref.
\cite{li2016scattering}.

\section{Conclusions and outlook \label{Conclusions}}

In this paper, we suggest a scheme for constructing probability thermodynamics
and quantum fields. In the scheme, we suppose that there is a fictional system
whose eigenvalues obey a probability distribution. Most characteristic
quantities in thermodynamics and quantum field theory are spectral functions,
such as various thermodynamics qualities and effective actions.

Our starting point is to analytically continue the heat kernel which is the
partition function in thermodynamics and the vacuum amplitude in quantum field
theory. We note here that the Wick-rotated heat kernel is a special case of
the analytic continuation of the heat kernel,%
\begin{equation}
K\left(  z\right)  =\sum_{n=-\infty}^{\infty}e^{\lambda_{n}z}, \label{Kz}%
\end{equation}
which is defined in the complex plane with a complex number $z$. By Eq.
(\ref{Kz}), we analytically continue the global heat kernel to the $z$-complex
plane. When $z$ is a pure imaginary number, the complex variable heat kernel,
$K\left(  z\right)  $, returns to Eq. (\ref{KtWick}). By the complex variable
heat kernel, $K\left(  z\right)  $, we can discuss the analyticity property of
the heat kernel on the complex $z$-plane.

In the scheme, we suppose that the eigenvalue obeys a probability distribution
without giving the operator. In further work, we can consider what an operator
corresponds to eigenvalue spectra obeying a certain given probability
distribution and find operators for various probability distributions. As a
rough example, the Gaussian distribution, $p\left(  x\right)  =\frac{1}%
{\sqrt{2\pi}\sigma}e^{-(x-\mu)^{2}/\left(  2\sigma^{2}\right)  }$, corresponds
to a Laplace operator $\Delta$. The negative Laplace operator $-\Delta$ is
lower bounded, corresponding to nonnegative random variables, and the positive
Laplace operator $+\Delta$ is lower unbounded, corresponding to negative
random variables.

The work of this paper bridges two fundamental theories: probability theory
and heat kernel theory. The heat kernel theory is important in both physics
and mathematics. In physics, the heat kernel theory plays an important role in
quantum field theory
\cite{vassilevich2003heat,mukhanov2007introduction,barvinsky1987beyond,barvinsky1990covariant,barvinsky1990covariant3,beauregard2015casimir,rajeev2011dispersion}%
. Scattering spectral theory is an important theory in quantum field theory
\cite{graham2009spectral,rahi2009scattering,weigel2018spectral,graham2014casimir}%
. The relation between heat kernels and scattering phases bridges the heat
kernel theory, the scattering spectral theory, and scattering theory
\cite{pang2012relation,li2015heat,liu2014scattering,weigel2018spectral,liu2022seeley,liu2023scattering}%
. This inspires us to relate the probability theory and the scattering theory,
which is an important issue in mathematical physics \cite{reed1979methods}.

As examples, we consider distributions with nonnegative random variables, such
as the Gamma distribution, the Beta distribution, the Chi-square distribution,
the Pareto distribution, the Generalized inverse Gaussian distribution, the
Tempered stable distribution, and the Log-normal distribution, and the
distributions with negative random variables, such as the Skewed normal
distribution, the Student's $t$-distribution, the Laplace distribution, the
Cauchy distribution, the $q$-Gaussian distribution, and the Intermediate distribution.

Probability theory is extensively applied across various disciplines,
including physics, biology, and social sciences. The Normal distribution, one
of the most significant statistical distributions, is ubiquitous across all
these fields. The Exponential distribution is relevant in the central spin
problem \cite{lindoy2018simple}, dislocation network
\cite{sills2018dislocation}, and the Anderson-like localization transition in
$SU(3)$ gauge theory {}\cite{kovacs2018localization}. The Beta distribution is
applied in the measurement of monoenergetic muon neutrino charged current
interactions \cite{aguilar2018first} and also in describing the collective
behavior of active particles aligning locally with neighbors {}%
\cite{laighleis2018minimal}. The Chi-square distribution is utilized in
neutrino experiments \cite{das2018determination}, microwave background
measurements{} \cite{aiola2015microwave}, and statistical inference{}
\cite{boes1974introduction,johnson1994continuous}. The Pareto distribution,
initially introduced in social science{} \cite{pareto1964cours}, is also found
in Calabi-Yau moduli spaces{} \cite{long2014heavy}, the Anderson localization
with a disordered medium{} \cite{asatryan2018anderson}, and the phenomenon of
cumulative inertia in intracellular transport{} \cite{fedotov2018memory}. The
Generalized Inverse Gaussian distribution, proposed for studying population
frequencies of species{} \cite{good1953population}, also referred to as the
Halphen type A distribution{} \cite{halphen1941nouveau}, is used to construct
the mixture of the Poisson distribution
\cite{sichel1974distribution,sichel1975distribution}. Barndorff-Nielsen has
identified that the mixture of the normal distribution and the generalized
inverse Gaussian distribution is the generalized hyperbolic distribution{}
\cite{barndorff1977exponentially,barndorff1978hyperbolic}. The generalized
inverse Gaussian distribution is found in fractional Brownian motion{}
\cite{wylomanska2016inverse} and the inverse Ising problem
\cite{donner2017inverse}. The tempered stable distribution, introduced in
\cite{tweedie1984index}, also known as the truncated Levy flight
\cite{koponen1995analytic} and the KoBoL distribution
\cite{boyarchenko2000option}, is generalized to the modified stable
distribution by Barndorff-Nielsen and Shephard \cite{barndorff2001normal}. It
is found in continuous-time random walks \cite{miyaguchi2013ergodic}. The
Log-normal distribution, a continuous probability distribution whose logarithm
is normally distributed, is found in global ocean models{}
\cite{pearson2018log}, particle physics {}\cite{nachman2015meta}, and the
issue of dark matter annihilation{} \cite{lisanti2018search}. The skewed
normal distribution, a generalization of the normal distribution with a shape
parameter accounting for skewness, was introduced by Azzalini{}
\cite{azzalini1985class}. It is found in gravitational wave problems
\cite{gair2015quantifying} and black hole problems \cite{torres2014formation}.
The Student's $t$-distribution, a symmetric and bell-shaped distribution used
to estimate the mean of a normally distributed population with a small sample
size and unknown population standard deviation, has heavier tails, meaning it
is more likely to produce values far from the mean. This is the primary
difference between the normal distribution and the Student's $t$-distribution.
The Laplace distribution, or double exponential distribution, can be
considered two exponential distributions joined back-to-back. It is a
symmetric distribution with fatter tails than the normal distribution and is
found in neuronal growth{} \cite{rizzo2013neuronal}, complex networks
{}\cite{skardal2014optimal}, and complex biological systems
\cite{stylianidou2018strong}. The Cauchy distribution, often used as an
example of a pathological distribution as it does not have finite moments of
order greater than one, but only has fractional absolute moments
\cite{johnson1994continuous}, is found in holographic dual problems{}
\cite{andrade2018coherence} and lasers \cite{raposo2015analytical}. The
$q$-Gaussian distribution, a generalization of the Gaussian distribution (as
the Tsallis entropy is a generalization of the standard Boltzmann-Gibbs
entropy or the Shannon entropy) \cite{tsallis2009nonadditive}, is an example
of the Tsallis distribution arising from the maximization of the Tsallis
entropy under appropriate constraints. The intermediate distribution, linking
the Gaussian and the Cauchy distribution{} \cite{liu2012intermediate}, reduces
to the Gaussian and the Cauchy distribution at specific parameter values. It
can be applied to spectral line broadening in laser theory and the stock
market return in quantitative finance {}\cite{liu2012intermediate}.

\appendix

\setcounter{section}{0}%
\setcounter{subsection}{0}%
\setcounter{subsubsection}{0}%
\setcounter{figure}{0}
\setcounter{specialtable}{0}
\setcounter{scheme}{0}
\setcounter{chart}{0}
\setcounter{boxenv}{0}
\setcounter{equation}{0}
\setcounter{theorem}{0}
\setcounter{lemma}{0}
\setcounter{corollary}{0}
\setcounter{proposition}{0} 
\setcounter{characterization}{0} 
\setcounter{property}{0} 
\setcounter{problem}{0} 
\setcounter{example}{0} 
\setcounter{examplesanddefinitions}{0} 
\setcounter{remark}{0} 
\setcounter{definition}{0} 
\setcounter{hypothesis}{0}
\setcounter{notation}{0}

\section[Distribution with nonnegative random variable]{Probability thermodynamics and probability quantum field of
distribution with nonnegative random variable \label{nonnegative}}

In this appendix, we list some examples for distribution with nonnegative
random variable, including the Exponential distribution, the Beta
distribution, the Chi-square distribution, the Pareto distribution, the
Generalized inverse Gaussian distribution, the Tempered stable distribution,
and the Log-normal distribution.

\subsection{Exponential distribution}

The probability density function
\cite{johnson1994continuous,ross2014introduction}%
\begin{equation}
p\left(  x\right)  =\lambda e^{-\lambda x},
\end{equation}
where $\lambda>0$ is the rate parameter.

The cumulative distribution function,%
\begin{equation}
P\left(  x\right)  =1-e^{-\lambda x}.
\end{equation}

\subsubsection{Thermodynamics}

The characteristic function%
\begin{equation}
f\left(  k\right)  =(1-ibk)^{-a}.
\end{equation}

The heat kernel%
\begin{equation}
K\left(  t\right)  =\frac{\lambda}{t+\lambda}.
\end{equation}
\ \ \ \ \ 

The partition function%
\begin{equation}
Z\left(  \beta\right)  =\frac{\lambda}{\beta+\lambda}.
\end{equation}

The free energy%
\begin{equation}
F=-\frac{1}{\beta}\ln\frac{\lambda}{\beta+\lambda}.
\end{equation}

The internal energy%
\begin{equation}
U=\frac{1}{\beta+\lambda}.
\end{equation}

The entropy%
\begin{equation}
S=\frac{\beta}{\beta+\lambda}+\ln\frac{\lambda}{\beta+\lambda}.
\end{equation}

The specific heat%
\begin{equation}
C_{V}=\frac{\beta^{2}}{(\beta+\lambda)^{2}}.
\end{equation}

\subsubsection{Quantum field}

The spectral zeta function%
\begin{equation}
\zeta\left(  s\right)  =\lambda^{s}\Gamma(1-s).
\end{equation}

The one-loop effective action
\begin{equation}
W=-\frac{\ln\lambda}{2}-\frac{\gamma_{\text{E}}}{2},
\end{equation}
where $\gamma_{\text{E}}\simeq0.577216$ is Euler's constant.

The vacuum energy%
\begin{equation}
E_{0}=\frac{1}{\lambda}.
\end{equation}

\subsection{Beta distribution}

The beta distribution
\begin{equation}
p\left(  x\right)  =\frac{(1-x)^{b-1}x^{a-1}}{B(a,b)},
\end{equation}
is defined on the interval $[0,1]$, where $a$ and $b$ are two positive shape
parameters \cite{johnson1994continuous}.

The cumulative distribution function,%
\begin{equation}
P\left(  x\right)  =I_{x}(a,b),
\end{equation}
where $I_{z}(a,b)=B\left(  z,a,b\right)  /B\left(  a,b\right)  $ is the
regularized incomplete beta function with $B\left(  z,a,b\right)  $ the
incomplete beta function \cite{abramowitz1964handbook}.

\subsubsection{Thermodynamics}

The characteristic function%
\begin{equation}
f\left(  k\right)  =\text{ }_{1}F_{1}(a;a+b;ik),
\end{equation}
where $_{1}F_{1}(a;b;z)$ is the Kummer confluent hypergeometric function.

The heat kernel%
\begin{equation}
K\left(  t\right)  =\text{ }_{1}F_{1}(a;a+b;-t).
\end{equation}

The partition function
\begin{equation}
Z\left(  \beta\right)  =\text{ }_{1}F_{1}(a;a+b;-\beta).
\end{equation}

The free energy%
\begin{equation}
F=-\frac{1}{\beta}\ln\,_{1}F_{1}(a;b+a;-\beta).
\end{equation}

The internal energy%
\begin{equation}
U=\frac{a\,_{1}F_{1}(a+1;b+a+1;-\beta)}{(a+b)\,_{1}F_{1}(a;b+a;-\beta)}.
\end{equation}

The entropy%
\begin{equation}
S=\frac{a\beta\,_{1}F_{1}(a+1;b+a+1;-\beta)}{(a+b)\,_{1}F_{1}(a;b+a;-\beta
)}+\ln\,_{1}F_{1}(a;b+a;-\beta).
\end{equation}

The specific heat%
\begin{equation}
C_{V}=\frac{a\beta^{2}}{(a+b)}\left[  \frac{(a+1)\,\,_{1}F_{1}%
(a+2;b+a+2;-\beta)}{\,_{1}F_{1}(a;b+a;-\beta){}\left(  a+b+1\right)  }%
-\frac{a\,_{1}F_{1}(a+1;b+a+1;-\beta){}^{2}}{(a+b)\,_{1}F_{1}(a;b+a;-\beta
){}^{2}}\right]  .
\end{equation}

\subsubsection{Quantum field}

The spectral zeta function%
\begin{equation}
\zeta\left(  s\right)  =\frac{\Gamma(b)\Gamma(a-s)}{B(a,b)\Gamma(b-s+a)}.
\end{equation}

The one-loop effective action
\begin{equation}
W=\frac{1}{2}\left[  \psi^{(0)}(a)-\psi^{(0)}(b+a)\right]  ,
\end{equation}
where $\psi^{(n)}\left(  z\right)  $ is the $n$-th derivative of the digamma function.

The vacuum energy%
\begin{equation}
E_{0}=\frac{a}{a+b}.
\end{equation}

\subsection{Chi-square distribution}

The chi-squared distribution is the distribution of a sum of the squares of
standard normal random variables. The chi-squared distribution with $\nu
$\ degrees of freedom \cite{johnson1994continuous}%
\begin{equation}
p\left(  x\right)  =\frac{1}{\Gamma\left(  \nu/2\right)  }2^{1-\nu/2}%
e^{-x^{2}/2}x^{\nu-1}.
\end{equation}

The cumulative distribution function%
\begin{equation}
P\left(  x\right)  =Q\left(  \nu/2,0,x/2\right)  ,
\end{equation}
where $Q\left(  a,z_{0},z_{1}\right)  $ is the generalized regularized
incomplete gamma function, defined as $\Gamma\left(  a,z_{0},z_{1}\right)
/\Gamma\left(  a\right)  $ with $\Gamma\left(  a,z_{0},z_{1}\right)  $ the
generalized incomplete gamma function \cite{olver2010nist}.

\subsubsection{Thermodynamics}

The characteristic function%
\begin{equation}
f\left(  k\right)  =\text{ }_{1}F_{1}\left(  \frac{\nu}{2};\frac{1}{2}%
;-\frac{k^{2}}{2}\right)  +\sqrt{2}\frac{ik}{\Gamma\left(  \nu/2\right)
}\Gamma\left(  \frac{\nu}{2}+\frac{1}{2}\right)  \,_{1}F_{1}\left(  \frac{\nu
}{2}+\frac{1}{2};\frac{3}{2};-\frac{k^{2}}{2}\right)  .
\end{equation}

The heat kernel%
\begin{equation}
K\left(  t\right)  =\text{ }_{1}F_{1}\left(  \frac{\nu}{2};\frac{1}{2}%
;\frac{t^{2}}{2}\right)  -t\sqrt{2}\frac{\Gamma\left(  \left(  \nu+1\right)
/2\right)  }{\Gamma\left(  \nu/2\right)  }\,_{1}F_{1}\left(  \frac{\nu+1}%
{2};\frac{3}{2};\frac{t^{2}}{2}\right)  .
\end{equation}

The partition function
\begin{equation}
Z\left(  \beta\right)  =\frac{1}{\sqrt{\pi}}\Gamma\left(  \frac{\nu+1}%
{2}\right)  U\left(  \frac{\nu}{2},\frac{1}{2},\frac{\beta^{2}}{2}\right)  .
\end{equation}
where $U\left(  a,b,z\right)  $ is the confluent hypergeometric function.

The free energy%
\begin{equation}
F=-\frac{1}{\beta}\ln\left(  \frac{1}{\sqrt{\pi}}\Gamma\left(  \frac{\nu+1}%
{2}\right)  U\left(  \frac{\nu}{2},\frac{1}{2},\frac{\beta^{2}}{2}\right)
\right)  .
\end{equation}

The internal energy%
\begin{align}
U  &  =\frac{2^{\nu-1}}{3\Gamma(\nu)U\left(  \nu/2,1/2,\beta^{2}/2\right)
}\left\{  \sqrt{2}\Gamma\left(  \frac{\nu+1}{2}\right)  \left[  (\nu
+1)\beta^{2}\text{ }_{1}F_{1}\left(  \frac{\nu+3}{2};\frac{5}{2};\frac
{\beta^{2}}{2}\right)  \right.  \right. \nonumber\\
&  \left.  \left.  +3\text{ }_{1}F_{1}\left(  \frac{\nu+1}{2};\frac{3}%
{2};\frac{\beta^{2}}{2}\right)  \right]  -6\beta\Gamma\left(  \frac{\nu}%
{2}+1\right)  \text{ }_{1}F_{1}\left(  \frac{\nu}{2}+1;\frac{3}{2};\frac
{\beta^{2}}{2}\right)  \right\}  .
\end{align}

The entropy%
\begin{align}
S  &  =\frac{\beta}{12\Gamma\left(  \nu/2\right)  \Gamma(\nu)U\left(
\nu/2,1/2,\beta^{2}/2\right)  }\nonumber\\
&  \times\left\{  \sqrt{2}\Gamma(\nu)\left\{  4\sqrt{\pi}\left[  \beta^{2}%
(\nu+1)\,_{1}F_{1}\left(  \frac{\nu+3}{2};\frac{5}{2};\frac{\beta^{2}}%
{2}\right)  +3\,_{1}F_{1}\left(  \frac{\nu+1}{2};\frac{3}{2};\frac{\beta^{2}%
}{2}\right)  \right]  \right\}  \right. \nonumber\\
&  \left.  -3\Gamma\left(  \frac{\nu}{2}\right)  U\left(  \frac{\nu+1}%
{2},\frac{3}{2},\frac{\beta^{2}}{2}\right)  \left[  \ln\pi-2\ln\left(
\Gamma\left(  \frac{\nu+1}{2}\right)  U\left(  \frac{\nu}{2},\frac{1}{2}%
,\frac{\beta^{2}}{2}\right)  \right)  \right]  \right. \nonumber\\
&  \left.  -3\beta2^{\nu+2}\Gamma\left(  \frac{\nu}{2}+1\right)  \Gamma\left(
\nu/2\right)  \,_{1}F_{1}\left(  \frac{\nu}{2}+1;\frac{3}{2};\frac{\beta^{2}%
}{2}\right)  \right\}  .
\end{align}

The specific heat%
\begin{align}
C_{V}  &  =-\frac{\beta^{2}}{45\left[  \Gamma\left(  \nu/2\right)  \,_{1}%
F_{1}\left(  \nu/2,1/2,\beta^{2}/2\right)  -\sqrt{2}\beta\Gamma\left(  \left(
\nu+1\right)  /2\right)  \,_{1}F_{1}\left(  \left(  \nu+1\right)
/2;3/2;\beta^{2}/2\right)  \right]  ^{2}}\nonumber\\
&  \times\left\{  5\left\{  \sqrt{2}\Gamma\left(  \frac{\nu+1}{2}\right)
\left[  \beta^{2}(\nu+1)\,_{1}F_{1}\left(  \frac{\nu+3}{2};\frac{5}{2}%
;\frac{\beta^{2}}{2}\right)  +3\,_{1}F_{1}\left(  \frac{\nu+1}{2};\frac{3}%
{2};\frac{\beta^{2}}{2}\right)  \right]  \right.  \right. \nonumber\\
&  \left.  \left.  -3\beta\nu\Gamma\left(  \frac{\nu}{2}\right)  \,_{1}%
F_{1}\left(  \frac{\nu}{2}+1;\frac{3}{2};\frac{\beta^{2}}{2}\right)  \right\}
^{2}\right.  \left.  +3\left[  \Gamma\left(  \frac{\nu}{2}\right)  \,_{1}%
F_{1}\left(  \frac{\nu}{2};\frac{1}{2};\frac{\beta^{2}}{2}\right)  -\sqrt
{2}\beta\Gamma\left(  \frac{\nu+1}{2}\right)  \,_{1}F_{1}\left(  \frac{\nu
+1}{2};\frac{3}{2};\frac{\beta^{2}}{2}\right)  \right]  \right. \nonumber\\
&  \times\left.  \left.  \left\{  \sqrt{2}\beta(\nu+1)\Gamma\left(  \frac
{\nu+1}{2}\right)  \left[  \beta^{2}(\nu+3)\,_{1}F_{1}\left(  \frac{\nu+5}%
{2};\frac{7}{2};\frac{\beta^{2}}{2}\right)  \right.  \right.  \right.  \right.
\nonumber\\
&  +\left.  15\,_{1}F_{1}\left(  \frac{\nu+3}{2};\frac{5}{2};\frac{\beta^{2}%
}{2}\right)  \right]  -5\nu\Gamma\left(  \frac{\nu}{2}\right)  \left[
\beta^{2}(\nu+2)_{1}F_{1}\left(  \frac{\nu}{2}+2;\frac{5}{2};\frac{\beta^{2}%
}{2}\right)  \right.  +\left.  \left.  \left.  3\,_{1}F_{1}\left(  \frac{\nu
}{2}+1;\frac{3}{2};\frac{\beta^{2}}{2}\right)  \right]  \right\}  \right\}  .
\end{align}

\subsubsection{Quantum field}

The spectral zeta function%
\begin{equation}
\zeta\left(  s\right)  =\frac{\Gamma\left(  \left(  \nu-s\right)  /2\right)
}{2^{s/2}\Gamma\left(  \nu/2\right)  }.
\end{equation}

The one-loop effective action
\begin{equation}
W=\frac{1}{2}\left(  \psi^{(0)}\left(  \frac{\nu}{2}\right)  +\ln2\right)  ,
\end{equation}
where $\psi^{(n)}\left(  z\right)  $ is the $n$-th derivative of the digamma function.

The vacuum energy%
\begin{equation}
E_{0}=\nu.
\end{equation}

\subsection{Pareto distribution}

The Pareto distribution \cite{johnson1994continuous}%
\begin{equation}
p\left(  x\right)  =b^{\alpha}x^{-1-\alpha}\alpha,
\end{equation}
where $b$ is the minimum value parameter and $\alpha$ is the shape parameter.

The cumulative distribution function%
\begin{equation}
P\left(  x\right)  =1-\left(  \frac{b}{x}\right)  ^{\alpha}.
\end{equation}

\subsubsection{Thermodynamics}

The characteristic function,%
\begin{equation}
f\left(  k\right)  =\alpha E_{\alpha+1}(-ibk),
\end{equation}
where $E_{n}(z)$ is the exponential integral function.

The heat kernel%
\begin{equation}
K\left(  t\right)  =\alpha E_{\alpha+1}(bt).
\end{equation}

The partition function
\begin{equation}
Z\left(  \beta\right)  =\alpha E_{\alpha+1}(b\beta).
\end{equation}

The free energy%
\begin{equation}
F=-\frac{1}{\beta}\ln\left(  \alpha E_{\alpha+1}(b\beta)\right)  .
\end{equation}

The internal energy%
\begin{equation}
U=\frac{bE_{\alpha}(b\beta)}{E_{\alpha+1}(b\beta)}.
\end{equation}

The entropy%
\begin{equation}
S=\frac{b\beta E_{\alpha}(b\beta)}{E_{\alpha+1}(b\beta)}+\ln\left(  \alpha
E_{\alpha+1}(b\beta)\right)  .
\end{equation}

The specific heat%
\begin{equation}
C_{V}=\frac{b^{2}\beta^{2}\left[  E_{\alpha-1}(b\beta)E_{\alpha+1}%
(b\beta)-E_{\alpha}(b\beta)^{2}\right]  }{E_{\alpha+1}(b\beta)^{2}}.
\end{equation}

\subsubsection{Quantum field}

The spectral zeta function%
\begin{equation}
\zeta\left(  s\right)  =\frac{\alpha}{s+\alpha}b^{-s}.
\end{equation}

The one-loop effective action
\begin{equation}
W=\frac{1}{2}\left(  \frac{1}{\alpha}+\ln b\right)  ,
\end{equation}
where $\psi^{(n)}\left(  z\right)  $ is the $n$-th derivative of the digamma function.

The vacuum energy%
\begin{equation}
E_{0}=\frac{\alpha}{\alpha-1}b.
\end{equation}

\subsection{Generalized inverse Gaussian distribution}

The generalized inverse Gaussian distribution \cite{johnson1994continuous}%
\begin{equation}
p\left(  x\right)  =\frac{\mu^{-\theta}x^{\theta-1}}{2K_{\theta}\left(
\lambda/\mu\right)  }\exp\left(  -\frac{\lambda}{2x}\left(  \frac{x^{2}}%
{\mu^{2}}+1\right)  \right)  ,
\end{equation}
where $\mu>0$ is the mean, $\lambda>0$ is the shape parameter, and $\theta$ is
a real number. $K_{n}(z)$\ is the modified Bessel function of the second kind
\cite{abramowitz1964handbook}. The generalized inverse Gaussian distribution
recovers the inverse Gaussian distribution when $\theta=-\frac{1}{2}$.

\subsubsection{Thermodynamics}

The characteristic function%
\begin{equation}
f\left(  k\right)  =\left(  1-\frac{2ik\mu^{2}}{\lambda}\right)  ^{-\theta
/2}\frac{K_{\theta}\left(  \left(  \lambda/\mu\right)  \sqrt{1-\frac
{2ik\mu^{2}}{\lambda}}\right)  }{K_{\theta}\left(  \lambda/\mu\right)  }.
\end{equation}

The heat kernel%
\begin{equation}
K\left(  t\right)  =\left(  \frac{2\mu^{2}t}{\lambda}+1\right)  ^{-\theta
/2}\frac{K_{\theta}\left(  \sqrt{\lambda\left(  2t\mu^{2}+\lambda\right)
}/\mu\right)  }{K_{\theta}\left(  \lambda/\mu\right)  }.
\end{equation}

The partition function
\begin{equation}
Z\left(  \beta\right)  =\frac{1}{K_{\theta}\left(  \lambda/\mu\right)
}\left(  \frac{\lambda}{2\beta\mu^{2}+\lambda}\right)  ^{\theta/2}K_{\theta
}\left(  \sigma\right)  .
\end{equation}

The free energy%
\begin{equation}
F=-\frac{1}{\beta}\ln\left(  \left(  \frac{\lambda}{2\beta\mu^{2}+\lambda
}\right)  ^{\theta/2}\frac{K_{\theta}\left(  \sigma\right)  }{K_{\theta
}\left(  \lambda/\mu\right)  }\right)  .
\end{equation}

The internal energy%
\begin{equation}
U=\frac{\mu}{K_{\theta}\left(  \sigma\right)  }\sqrt{\frac{\lambda}{2\beta
\mu^{2}+\lambda}}K_{\theta+1}\left(  \sigma\right)  .
\end{equation}

The entropy%
\begin{equation}
S=\frac{\beta\lambda K_{\theta+1}\left(  \sigma\right)  }{\sigma K_{\theta
}\left(  \sigma\right)  }+\ln\left(  \left(  \frac{\lambda}{2\beta\mu
^{2}+\lambda}\right)  ^{\theta/2}\frac{K_{\theta}\left(  \sigma\right)
}{K_{\theta}\left(  \lambda/\mu\right)  }\right)  .
\end{equation}

The specific heat%
\begin{align}
C_{V}  &  =\frac{\beta^{2}\lambda^{2}}{4\sigma^{4}}\left[  2\sigma
\frac{K_{\theta+1}\left(  \sigma\right)  }{K_{\theta}\left(  \sigma\right)
}+\lambda\frac{\left(  2\beta\mu^{2}-\lambda\right)  }{\mu^{2}}\frac
{K_{\theta+1}\left(  \sigma\right)  ^{2}}{K_{\theta}\left(  \sigma\right)
^{2}}+\sigma^{2}\frac{K_{\theta+2}\left(  \sigma\right)  }{K_{\theta}\left(
\sigma\right)  }-\sigma^{2}\frac{K_{\theta-1}\left(  \sigma\right)  ^{2}%
}{K_{\theta}\left(  \sigma\right)  ^{2}}\right. \nonumber\\
&  \left.  +2\sigma\frac{K_{\theta-1}\left(  \sigma\right)  }{K_{\theta
}\left(  \sigma\right)  }+\sigma^{2}\frac{K_{\theta-2}\left(  \sigma\right)
}{K_{\theta}\left(  \sigma\right)  }-2\sigma^{2}\frac{K_{\theta-1}\left(
\sigma\right)  K_{\theta+1}\left(  \sigma\right)  }{K_{\theta}\left(
\sigma\right)  ^{2}}+\frac{2\lambda^{2}}{\mu^{2}}+8\theta+4\beta
\lambda\right]  ,
\end{align}
where $\sigma=\sqrt{\lambda\left(  2\beta\mu^{2}+\lambda\right)  }/\mu$.

\subsubsection{Quantum field}

The spectral zeta function%
\begin{equation}
\zeta\left(  s\right)  =\frac{\mu^{-s}K_{s-\theta}\left(  \lambda/\mu\right)
}{K_{\theta}\left(  \lambda/\mu\right)  }.
\end{equation}

The one-loop effective action
\begin{equation}
W=\frac{1}{2}\left(  \ln\mu+\frac{1}{K_{\theta}\left(  \lambda/\mu\right)
}\frac{\partial K_{-\theta}\left(  \lambda/\mu\right)  }{\partial\theta
}\right)  ,
\end{equation}
where $K_{n}\left(  z\right)  $ gives the modified Bessel function of the
second kind.

The vacuum energy%
\begin{equation}
E_{0}=\mu\frac{K_{\theta+1}\left(  \lambda/\mu\right)  }{K_{\theta}\left(
\lambda/\mu\right)  }.
\end{equation}

\subsection{Tempered stable distribution}

The tempered stable distribution \cite{schoutens2003levy}%
\begin{equation}
p\left(  x;\kappa,\alpha,b\right)  =\frac{e^{\alpha b}e^{-\frac{1}%
{2}b^{1/\kappa}x}}{2\pi\alpha^{1/\kappa}}\sum_{k=1}^{\infty}\left(  -1\right)
^{k-1}\sin\left(  k\pi\kappa\right)  \frac{\Gamma\left(  k\kappa+1\right)
}{k!}2^{k\kappa+1}\left(  \frac{x}{\alpha^{1/\kappa}}\right)  ^{-k\kappa-1},
\end{equation}
where $\alpha>0$, $b\geq0$, and $0<\kappa<1$.

\subsubsection{Thermodynamics}

The characteristic function%
\begin{equation}
f\left(  k\right)  =\exp\left(  ab-a\left(  b^{1/\kappa}-2ik\right)  ^{\kappa
}\right)  .
\end{equation}
For $\kappa=1/2$ the tempered stable distribution reduces to the inverse
Gaussian distribution. For the limiting case $\kappa\rightarrow0$, we obtain
the Gamma\ distribution.

The heat kernel%
\begin{equation}
K\left(  t\right)  =\exp\left(  ab-a\left(  b^{1/\kappa}+2t\right)  ^{\kappa
}\right)  .
\end{equation}

The partition function
\begin{equation}
Z\left(  \beta\right)  =\exp\left(  ab-a\left(  b^{1/\kappa}+2\beta\right)
^{\kappa}\right)  .
\end{equation}

The free energy%
\begin{equation}
F=-\frac{a}{\beta}\left[  b-\left(  b^{1/\kappa}+2\beta\right)  ^{\kappa
}\right]  .
\end{equation}

The internal energy%
\begin{equation}
U=2a\left(  b^{1/\kappa}+2\beta\right)  ^{-1+\kappa}\kappa.
\end{equation}

The entropy%
\begin{equation}
S=2a\beta\left(  b^{1/\kappa}+2\beta\right)  ^{-1+\kappa}\kappa+a\left[
b-\left(  b^{1/\kappa}+2\beta\right)  ^{\kappa}\right]  .
\end{equation}

The specific heat%
\begin{equation}
C_{V}=-4a\beta^{2}\left(  b^{1/\kappa}+2\beta\right)  ^{-2+\kappa}%
(\kappa-1)\kappa.
\end{equation}

\subsubsection{Quantum field}

The spectral zeta function%
\begin{equation}
\zeta\left(  s\right)  =-2^{-s}e^{ab}\sin\kappa\sum_{k=1}^{\infty}%
(-1)^{k}\frac{\Gamma(k\kappa+1)\Gamma(-s-k\kappa)}{\left(  k-1\right)  !}%
a^{k}b^{k+s/\kappa}.
\end{equation}

The one-loop effective action
\begin{equation}
W=-\frac{1}{2}\sum_{k=1}^{\infty}\frac{\pi(-a)^{k}b^{k}e^{ab}}{\Gamma
(k)}\left[  \ln b\sin(\csc(\pi\kappa k))-\ln2\sin\kappa\csc(\pi\kappa
k)-\sin\kappa\csc(\pi\kappa k)\psi^{(0)}(-k\kappa)\right]  .
\end{equation}

The vacuum energy%
\begin{equation}
E_{0}=2a\kappa b^{\left(  \kappa-1\right)  /\kappa}.
\end{equation}

\subsection{Log-normal distribution}

Thelog-normal distribution \cite{johnson1994continuous}%
\begin{equation}
p\left(  x\right)  =\frac{1}{\sqrt{2\pi}x\sigma}\exp\left(  -\frac{(-\mu+\ln
x)^{2}}{2\sigma^{2}}\right)  .
\end{equation}
The logarithm of a log-normal distribution $X$ with parameters $%
%TCIMACRO{\U{3bc} }%
%BeginExpansion
\mu
%EndExpansion
$\ and $\sigma$ is\ normally distributed. $\ln X$\ has a mean $%
%TCIMACRO{\U{3bc} }%
%BeginExpansion
\mu
%EndExpansion
$\ and standard deviation $\sigma$. Then we can write $X$\ as%
\begin{equation}
X=\exp\left(  \mu+\sigma Z\right)
\end{equation}
with $Z$\ a standard normal variable.

The cumulative distribution function%
\begin{equation}
P\left(  x\right)  =\frac{1}{2}\operatorname{erfc}\left(  \frac{\mu-\ln
x}{\sqrt{2}\sigma}\right)  ,
\end{equation}
where $\operatorname{erfc}\left(  x\right)  $ is the complementary error
function given by $\operatorname{erfc}\left(  z\right)  =1-\operatorname{erf}%
(z)$ with $\operatorname{erf}(z)$ the integral of the Gaussian distribution:
$\frac{2}{\sqrt{\pi}}\int_{0}^{z}e^{-t^{2}}dt$.

\subsubsection{Thermodynamics}

The characteristic function%
\begin{equation}
f\left(  k\right)  =\sum_{n=0}^{\infty}\frac{(ik)^{n}}{n!}\exp\left(
n\mu+\frac{n^{2}\sigma^{2}}{2}\right)  .
\end{equation}

The heat kernel%
\begin{equation}
K\left(  t\right)  =\sum_{n=0}^{\infty}\frac{(-t)^{n}}{n!}\exp\left(
n\mu+\frac{n^{2}\sigma^{2}}{2}\right)  .
\end{equation}

The partition function
\begin{equation}
Z\left(  \beta\right)  =\sum_{n=0}^{\infty}\frac{(-\beta)^{n}}{n!}\exp\left(
n\mu+\frac{n^{2}\sigma^{2}}{2}\right)  .
\end{equation}

The free energy%
\begin{equation}
F=-\frac{1}{\beta}\ln\sum_{n=0}^{\infty}\frac{(-\beta)^{n}}{n!}\exp\left(
n\mu+\frac{n^{2}\sigma^{2}}{2}\right)  .
\end{equation}

The internal energy%
\begin{equation}
U=-\frac{\sum_{n=0}^{\infty}-\frac{n(-\beta)^{-1+n}}{n!}\exp\left(  n\mu
+\frac{n^{2}\sigma^{2}}{2}\right)  }{\sum_{n=0}^{\infty}\frac{(-\beta)^{n}%
}{n!}\exp\left(  n\mu+\frac{n^{2}\sigma^{2}}{2}\right)  }.
\end{equation}

The entropy%
\begin{equation}
S=\ln\sum_{n=0}^{\infty}\frac{(-\beta)^{n}}{n!}\exp\left(  n\mu+\frac
{n^{2}\sigma^{2}}{2}\right)  -\frac{\sum_{n=0}^{\infty}-\frac{n(-\beta)^{n}%
}{n!}\exp\left(  n\mu+\frac{n^{2}\sigma^{2}}{2}\right)  }{\sum_{n=0}^{\infty
}\frac{(-\beta)^{n}}{n!}\exp\left(  n\mu+\frac{n^{2}\sigma^{2}}{2}\right)  }.
\end{equation}

The specific heat%
\begin{align}
C_{V}  &  =\frac{\beta^{2}}{\left[  \sum_{n=0}^{\infty}\frac{(-1)^{n}\beta
^{n}}{n!}\exp\left(  n\mu+\frac{n^{2}\sigma^{2}}{2}\right)  \right]  {}^{2}%
}\nonumber\\
&  \times\left\{  \sum_{n=0}^{\infty}\frac{\left(  -\beta\right)  ^{n-2}%
}{(n-2)!}e^{n^{2}\sigma^{2}/2+\mu n}\sum_{n=0}^{\infty}\frac{\left(
-\beta\right)  ^{n}}{n!}e^{n^{2}\sigma^{2}/2+\mu n}-\left[  \sum_{n=0}%
^{\infty}-\frac{(-\beta)^{n-1}}{(n-1)!}e^{n^{2}\sigma^{2}/2+\mu n}\right]
^{2}\right\}  .
\end{align}

\subsubsection{Quantum field}

The spectral zeta function%
\begin{equation}
\zeta\left(  s\right)  =\exp\left(  \frac{1}{2}s\left(  s\sigma^{2}%
-2\mu\right)  \right)  .
\end{equation}

The one-loop effective action
\begin{equation}
W=\frac{\mu}{2}.
\end{equation}

The vacuum energy%
\begin{equation}
E_{0}=\exp\left(  \mu+\frac{\sigma^{2}}{2}\right)  .
\end{equation}

\section[Distribution with negative random variable]{Probability thermodynamics and probability quantum field of
distribution with negative random variable \label{negative}}

In this appendix, we list some examples for distribution with negative random
variable, including the Skewed normal distribution, the Student's
$t$-distribution, the Laplace distribution, the Cauchy distribution, the
$q$-Gaussian distribution, and the intermediate distribution.

\subsection{Skewed normal distribution}

The Skewed Gaussian distribution \cite{azzalini1985class}%
\begin{equation}
p\left(  x\right)  =\frac{1}{\sqrt{2\pi}\sigma}e^{-(x-\mu)^{2}/\left(
2\sigma^{2}\right)  }\operatorname{erfc}\left(  -\frac{\alpha(x-\mu)}{\sqrt
{2}\sigma}\right)  ,
\end{equation}
where $\alpha$ is the shape parameter, $\mu$\ is the location parameter, and
$\sigma$ is the scale parameter. The Gaussian\ distribution is recovered when
$\alpha=0$. \ 

The cumulative distribution function%
\begin{equation}
P\left(  x\right)  =\frac{1}{2}\operatorname{erfc}\left(  -\frac{x-\mu}%
{\sqrt{2}\sigma}\right)  -2T\left(  \frac{x-\mu}{\sigma},\alpha\right)  ,
\end{equation}
where $T\left[  x,\alpha\right]  $ is Owen's $T$ function given by $T\left[
x,\alpha\right]  =\frac{1}{2\pi}\int_{0}^{\alpha}\frac{e^{-x^{2}\left(
1+t^{2}\right)  /2}}{1+t^{2}}dt$ for real $\alpha$ \cite{owen1956tables}.

\subsubsection{Thermodynamics}

The characteristic function%
\begin{equation}
f\left(  k\right)  =e^{-\frac{1}{2}k^{2}\sigma^{2}+ik\mu}\operatorname{erfc}%
\left(  \frac{i\alpha k\sigma}{\sqrt{2}\sqrt{\alpha^{2}+1}}\right)  .
\end{equation}

The heat kernel%
\begin{equation}
K\left(  t\right)  =e^{\frac{1}{2}t\left(  \sigma^{2}t-2\mu\right)
}\operatorname{erfc}\left(  \frac{\alpha\sigma t}{\sqrt{2}\sqrt{\alpha^{2}+1}%
}\right)  .
\end{equation}

The partition function
\begin{equation}
Z\left(  \beta\right)  =e^{\frac{1}{2}\beta\left(  \beta\sigma^{2}%
-2\mu\right)  }\operatorname{erfc}\left(  \frac{\alpha\sigma\beta}%
{\sqrt{2\alpha^{2}+2}}\right)  .
\end{equation}

The free energy%
\begin{equation}
F=-\frac{1}{2}\left(  \beta\sigma^{2}-2\mu\right)  +\frac{1}{\beta}%
\ln\operatorname{erfc}\left(  \frac{\alpha\sigma\beta}{\sqrt{2\alpha^{2}+2}%
}\right)  .
\end{equation}

The internal energy%
\begin{equation}
U=\mu-\beta\sigma^{2}+\frac{2\sigma\alpha}{\sqrt{\pi}\sqrt{2\alpha^{2}+2}%
}\frac{\exp\left(  -\alpha^{2}\sigma^{2}\beta^{2}/\left(  2\alpha
^{2}+2\right)  \right)  }{\operatorname{erfc}\left(  \beta\sigma\alpha
/\sqrt{2\alpha^{2}+2}\right)  }.
\end{equation}

The entropy%
\begin{equation}
S=\frac{2\alpha\beta\sigma\exp\left(  -\alpha^{2}\sigma^{2}\beta^{2}/\left(
2\alpha^{2}+2\right)  \right)  }{\sqrt{\pi}\sqrt{2\alpha^{2}+2}%
\operatorname{erfc}\left(  \beta\sigma\alpha/\sqrt{2\alpha^{2}+2}\right)
}+\ln\left(  \operatorname{erfc}\left(  \frac{\alpha\beta\sigma}{\sqrt
{2\alpha^{2}+2}}\right)  \right)  -\frac{1}{2}\beta^{2}\sigma^{2}.
\end{equation}

The specific heat%
\begin{equation}
C_{V}=\frac{\alpha^{2}\beta^{2}\sigma^{2}e^{-\alpha^{2}\beta^{2}\sigma
^{2}/\alpha^{2}+1}\left[  \sqrt{2\pi}\alpha\beta\sigma e^{\alpha^{2}\beta
^{2}\sigma^{2}/2\alpha^{2}+2}\operatorname{erfc}\left(  \beta\sigma
\alpha/\sqrt{2\alpha^{2}+2}\right)  -2\sqrt{\alpha^{2}+1}\right]  }{\pi\left(
\alpha^{2}+1\right)  ^{3/2}\operatorname{erfc}\left(  \beta\sigma\alpha
/\sqrt{2\alpha^{2}+2}\right)  ^{2}}+\beta^{2}\sigma^{2}.
\end{equation}

\subsubsection{Quantum field}

The spectral zeta function%
\begin{equation}
\zeta\left(  s\right)  =\frac{1}{\sqrt{2\pi}\sigma}\int_{-\infty}^{\infty
}e^{-(x-\mu)^{2}/\left(  2\sigma^{2}\right)  }\operatorname{erfc}\left(
-\frac{\alpha(x-\mu)}{\sqrt{2}\sigma}\right)  x^{-s}dx. \label{SkewedZeta}%
\end{equation}

The one-loop effective action
\begin{equation}
W=\frac{1}{2\sqrt{2\pi}\sigma}\int_{-\infty}^{\infty}e^{-(x-\mu)^{2}/\left(
2\sigma^{2}\right)  }\operatorname{erfc}\left(  -\frac{\alpha(x-\mu)}{\sqrt
{2}\sigma}\right)  \ln xdx. \label{SkewedWs}%
\end{equation}

The vacuum energy%
\begin{equation}
E_{0}=\mu+\sqrt{\frac{2}{\pi}}\frac{\alpha\sigma}{\sqrt{\alpha^{2}+1}}.
\end{equation}

When $\mu=0$, the integral in Eqs. (\ref{SkewedZeta}) and (\ref{SkewedWs}) can
be carried out,%
\begin{align}
\zeta\left(  s\right)   &  =\frac{1}{\sqrt{\pi}}\sigma^{-s}\left\{
2^{s/2-1}\left[  (-1)^{s-1}+1\right]  \alpha^{s-1}\Gamma(1-s)\,_{2}\tilde
{F}_{1}\left(  \frac{1-s}{2},1-\frac{s}{2};\frac{3-s}{2};-\frac{1}{\alpha^{2}%
}\right)  \right. \nonumber\\
&  +\left.  2^{-s/2}(-1)^{s}\Gamma\left(  \frac{1}{2}-\frac{s}{2}\right)
\right\}  .
\end{align}
and
\begin{equation}
W=\frac{1}{4}\left(  2i\cot^{-1}\alpha+\ln2\sigma^{2}+\psi^{(0)}\left(
\frac{1}{2}\right)  \right)  .
\end{equation}

The vacuum energy%
\begin{equation}
E_{0}=\sqrt{\frac{2}{\pi}}\frac{\alpha\sigma}{\sqrt{\alpha^{2}+1}}.
\end{equation}

\subsection{Student's $t$-distribution}

The Student's $t$-distribution \cite{johnson1995continuous}
\begin{equation}
p\left(  x\right)  =\frac{1}{\sqrt{\nu}\sigma B\left(  \frac{\nu}{2},\frac
{1}{2}\right)  }\left[  \frac{\nu}{\nu+(x-\mu)^{2}/\sigma^{2}}\right]
^{\left(  \nu+1\right)  /2},
\end{equation}
where $\mu$ is location parameter, $\sigma$ scale parameter, and $\nu$ degrees
of freedom. For $\nu\rightarrow\infty$, the Student's $t$-distribution goes to
Gaussian distribution. For $\nu=1$, Student's $t$-distribution goes to Cauchy distribution.

The cumulative distribution function%
\begin{equation}
P\left(  x\right)  =\left\{
\begin{array}
[c]{c}%
\displaystyle\frac{1}{2}I_{\frac{\nu\sigma^{2}}{(x-\mu)^{2}+\nu\sigma^{2}}%
}\left(  \frac{\nu}{2},\frac{1}{2}\right)  ,\text{ \ \ }x\leq\mu,\\
\displaystyle\frac{1}{2}\left[  I_{\frac{(x-\mu)^{2}}{(x-\mu)^{2}+\nu
\sigma^{2}}}\left(  \frac{1}{2},\frac{\nu}{2}\right)  +1\right]  ,\text{
\ }x>\mu,
\end{array}
\right.
\end{equation}
where $B\left(  a,b\right)  $ is the Euler beta function, $K_{n}(z)$ is the
modified Bessel function of the second kind, and $I_{z}(a,b)$ is regularized
incomplete beta function given by $I_{z}(a,b)=B\left(  z,a,b\right)  /B\left(
a,b\right)  $ \cite{abramowitz1964handbook}.

\subsubsection{Thermodynamics}

The characteristic function%
\begin{equation}
f\left(  k\right)  =\frac{2^{1-\nu/2}\nu^{\nu/4}\sigma^{\nu/2}}{\Gamma\left(
\nu/2\right)  }e^{ik\mu}|k|^{\nu/2}K_{\nu/2}\left(  \sqrt{\nu}\sigma
|k|\right)  .
\end{equation}

The heat kernel%
\begin{equation}
K\left(  t\right)  =\frac{2^{1-\nu/2}\nu^{\nu/4}}{\Gamma\left(  \nu/2\right)
}e^{-t\mu}\left(  \sigma t\right)  ^{\nu/2}K_{\nu/2}\left(  \sqrt{\nu}\sigma
t\right)  .
\end{equation}

The partition function
\begin{equation}
Z\left(  \beta\right)  =\frac{2^{1-\nu/2}\nu^{\nu/4}}{\Gamma\left(
\nu/2\right)  }e^{-\beta\mu}(\sigma\beta)^{\nu/2}K_{\nu/2}\left(  \beta
\sqrt{\nu}\sigma\right)  .
\end{equation}

The free energy%
\begin{equation}
F=-\frac{1}{\beta}\ln\frac{2^{1-\nu/2}\nu^{\nu/4}e^{-\beta\mu}(\beta
\sigma)^{\nu/2}K_{\nu/2}\left(  \beta\sqrt{\nu}\sigma\right)  }{\Gamma\left(
\nu/2\right)  }.
\end{equation}

The internal energy%
\begin{equation}
U=\frac{1}{2}\left\{  \sqrt{\nu}\sigma\frac{K_{\nu/2-1}\left(  \beta\sqrt{\nu
}\sigma\right)  +K_{\nu/2+1}\left(  \beta\sqrt{\nu}\sigma\right)  }{K_{\nu
/2}\left(  \beta\sqrt{\nu}\sigma\right)  }-\frac{\nu}{\beta}+2\mu\right\}  .
\end{equation}

The entropy%
\begin{equation}
S=\frac{1}{2}\left[  \ln\frac{4\nu^{\nu/2}e^{-2\beta\mu}(\beta\sigma)^{\nu
}K_{\nu/2}\left(  \beta\sqrt{\nu}\sigma\right)  {}^{2}}{\Gamma\left(
\nu/2\right)  ^{2}}+2\beta\mu+\frac{2\beta\sqrt{\nu}\sigma K_{\nu/2-1}\left(
\beta\sqrt{\nu}\sigma\right)  }{K_{\nu/2}\left(  \beta\sqrt{\nu}\sigma\right)
}-\nu\ln2\right]  .
\end{equation}

The specific heat%
\begin{align}
C_{V}  &  =\frac{-\beta^{2}\nu\sigma^{2}}{4K_{\nu/2}\left(  \beta\sqrt{\nu
}\sigma\right)  {}^{2}}\left\{  \left[  K_{\nu/2-1}\left(  \beta\sqrt{\nu
}\sigma\right)  +K_{\nu/2+1}\left(  \beta\sqrt{\nu}\sigma\right)  \right]
{}^{2}\right. \nonumber\\
&  -\left.  K_{\nu/2}\left(  \beta\sqrt{\nu}\sigma\right)  \left[  K_{\nu
/2-2}\left(  \beta\sqrt{\nu}\sigma\right)  +2K_{\nu/2}\left(  \beta\sqrt{\nu
}\sigma\right)  +K_{\nu/2+2}\left(  \beta\sqrt{\nu}\sigma\right)  \right]
\right\}  -\frac{\nu}{2}.
\end{align}

\subsubsection{Quantum field}

The spectral zeta function%
\begin{equation}
\zeta\left(  s\right)  =\frac{\sqrt{\pi}(-1)^{\nu/4}e^{i\pi(\nu+2s)/4}}%
{2^{\nu+s-1}}\left(  \frac{i}{\mu}\right)  ^{s}\frac{\Gamma(s+\nu)}%
{\Gamma\left(  \nu/2\right)  }\,_{2}\tilde{F}_{1}\left(  \frac{s}{2}%
,\frac{s+1}{2};\frac{1}{2}(2s+\nu+1);\frac{\nu\sigma^{2}}{\mu^{2}}+1\right)  ,
\end{equation}
where $_{2}\tilde{F}_{1}$ $\left(  a,b;c;d\right)  $ is the regularized
hypergeometric function and $_{2}\tilde{F}_{1}$ $\left(  a,b;c;d\right)  =$\\
$_{2}F_{1}\left(  a,b;c;d\right)  /\Gamma\left(  c\right)  $.

The one-loop effective action
\begin{align}
W  &  =\frac{(-1)^{\nu}}{4}\left[  \ln4\mu^{2}-2\psi^{(0)}(\nu)-2i\pi
-\Gamma\left(  \frac{\nu+1}{2}\right)  \left(  2_{2}\tilde{F}_{1}%
^{(0,0,1,0)}\left(  0,\frac{1}{2},\frac{\nu+1}{2},\frac{\nu\sigma^{2}}{\mu
^{2}}+1\right)  \right.  \right. \nonumber\\
&  \left.  +\left.  _{2}\tilde{F}_{1}^{(0,1,0,0)}\left(  0,\frac{1}{2}%
,\frac{\nu+1}{2},\frac{\nu\sigma^{2}}{\mu^{2}}+1\right)  +\text{ }_{2}%
\tilde{F}_{1}^{(1,0,0,0)}\left(  0,\frac{1}{2},\frac{\nu+1}{2},\frac{\nu
\sigma^{2}}{\mu^{2}}+1\right)  \right)  \right]  .
\end{align}

The vacuum energy%
\begin{equation}
E_{0}=\mu.
\end{equation}

\subsection{Laplace distribution}

The Laplace distribution \cite{johnson1995continuous,kotz2001laplace}%

\begin{equation}
p\left(  x\right)  =\frac{1}{2}e^{-\lambda\left\vert x-\mu\right\vert }%
\lambda,
\end{equation}
where $\mu$ is the mean and $\lambda$ is the scale parameter.

The cumulative distribution function%
\begin{equation}
P\left(  x\right)  =\left\{
\begin{array}
[c]{c}%
\displaystyle1-\frac{1}{2}e^{-\lambda\left(  x-\mu\right)  },\text{ \ }%
x\geq\mu\\
\displaystyle\frac{1}{2}e^{\lambda\left(  x-\mu\right)  },\text{ \ \ }x<\mu
\end{array}
\right.  .
\end{equation}

\subsubsection{Thermodynamics}

The characteristic function%
\begin{equation}
f\left(  k\right)  =\frac{\lambda^{2}}{k^{2}+\lambda^{2}}e^{ik\mu}.
\end{equation}

The heat kernel%
\begin{equation}
K\left(  t\right)  =\frac{\lambda^{2}}{\lambda^{2}-t^{2}}e^{-t\mu}.
\end{equation}

The partition function
\begin{equation}
Z\left(  \beta\right)  =\frac{\lambda^{2}}{\lambda^{2}-\beta^{2}}e^{-\beta\mu
}.
\end{equation}

The free energy%
\begin{equation}
F=-\frac{1}{\beta}\ln\frac{\lambda^{2}}{\lambda^{2}-\beta^{2}}e^{-\beta\mu}.
\end{equation}

The internal energy%
\begin{equation}
U=\frac{2\beta}{\beta^{2}-\lambda^{2}}+\mu.
\end{equation}

The entropy%
\begin{equation}
S=\left(  \beta\mu+1-\frac{2\beta^{2}}{\lambda^{2}-\beta^{2}}\right)  \left(
-\beta\mu+\ln\frac{\lambda^{2}}{\lambda^{2}-\beta^{2}}\right)  .
\end{equation}

The specific heat%
\begin{equation}
C_{V}=\frac{2\beta^{2}\left(  \beta^{2}+\lambda^{2}\right)  }{\left(
\beta^{2}-\lambda^{2}\right)  ^{2}}.
\end{equation}

\subsubsection{Quantum field}

The spectral zeta function%
\begin{equation}
\zeta\left(  s\right)  =\frac{1}{2}e^{-\lambda\mu}\mu^{-s}\left(  -\lambda\mu
E_{s}(-\lambda\mu)+e^{2\lambda\mu}(\lambda\mu)^{s}\Gamma(1-s,\lambda
\mu)-2i\sin(\pi s)\Gamma(1-s)(\lambda\mu)^{s}\right)  ,
\end{equation}
where $E_{n}(z)$ gives the exponential integral function

The one-loop effective action
\begin{equation}
W=\frac{1}{4}e^{-\lambda\mu}\left(  2e^{\lambda\mu}\ln\mu+\Gamma(0,-\lambda
\mu)+e^{2\lambda\mu}\Gamma(0,\lambda\mu)+2i\pi\right)  .
\end{equation}

The vacuum energy%
\begin{equation}
E_{0}=\mu.
\end{equation}

\subsection{$q$-Gaussian distribution}

The probability density function of the $q$-Gaussian distribution is
\cite{umarov2008q}%
\begin{equation}
p\left(  x\right)  =\frac{\sqrt{b}}{C_{q}}e_{q}\left(  -bx^{2}\right)  ,
\end{equation}
where $e_{q}\left(  x\right)  =\left[  (1-q)x+1\right]  ^{\frac{1}{1-q}}$ is
the $q$-exponential, $q$ is deformation parameter, $b$ is positive real
number, and the normalization factor $C_{q}$ is given by
\begin{equation}
C_{q}=\left\{
\begin{array}
[c]{c}%
\displaystyle\frac{2\sqrt{\pi}\Gamma\left(  \frac{1}{1-q}\right)  }{\left(
3-q\right)  \sqrt{1-q}\Gamma\left(  \frac{3-q}{2\left(  1-q\right)  }\right)
},\text{ \ }-\infty<q<1,\\
\sqrt{\pi},\text{ \ \ \ \ }q=1,\\
\displaystyle\frac{\sqrt{\pi}\Gamma\left(  \frac{3-q}{2(q-1)}\right)  }%
{\sqrt{q-1}\Gamma\left(  \frac{1}{q-1}\right)  },\text{ \ }1<q<3,
\end{array}
\right.  ,
\end{equation}
where $x\in\left(  -\infty,+\infty\right)  $ for $1\leq q<3$ and $x\in\left(
-\frac{1}{\sqrt{b\left(  1-q\right)  }},+\frac{1}{\sqrt{b\left(  1-q\right)
}}\right)  $ for $q<1$ with $b$ the scale parameter and $q$ the deformation
parameter. For $q=1$,\quad the $q$-Gaussian distribution reduces to the
Gaussian distribution. For $q=2$ and $\sigma=\sqrt{\frac{1}{b}}$, the
$q$-Gaussian distribution reduces to the Cauchy distribution.

The cumulative distribution function, taking $1<q<3$ as an example, by Eq.
(\ref{CPF}), reads%
\begin{equation}
P\left(  x\right)  =\left\{
\begin{array}
[c]{c}%
\displaystyle\frac{x\sqrt{b(q-1)}\Gamma\left(  \frac{1}{q-1}\right)  \,}%
{\sqrt{\pi}\Gamma\left(  \frac{1}{q-1}-\frac{1}{2}\right)  }\text{ }_{2}%
F_{1}\left(  \frac{1}{2},\frac{1}{q-1};\frac{3}{2};-b(q-1)x^{2}\right)
+\frac{1}{2},\text{ \ }1<q<3,\\
\displaystyle\frac{x\sqrt{b-bq}\Gamma\left(  \frac{3}{2}+\frac{1}{1-q}\right)
}{\sqrt{\pi}\Gamma\left(  1+\frac{1}{1-q}\right)  }\,_{2}F_{1}\left(  \frac
{1}{2},\frac{1}{q-1};\frac{3}{2};-b(q-1)x^{2}\right)  +\frac{1}{2},\text{
\ }q<1,\\
\displaystyle\frac{1}{2}\left(  \operatorname{erf}\left(  \sqrt{b}x\right)
+1\right)  ,\text{ \ }q=1,
\end{array}
\right.  ,
\end{equation}
where $\,_{2}F_{1}\left(  a,b;c;z\right)  $ is the hypergeometric function and
$\operatorname{erf}\left(  z\right)  $ is the error function.

\subsubsection{Thermodynamics}

The characteristic function%
\[
f\left(  k\right)  =\frac{2^{\frac{1}{1-q}+\frac{3}{2}}\left[  b(q-1)/|k|^{2}%
\right]  ^{\frac{q+1}{4-4q}+\frac{1}{2}}}{\Gamma\left(  \frac{1}{q-1}-\frac
{1}{2}\right)  }K_{\frac{1}{2}+\frac{1}{1-q}}\left(  \frac{|k|}{\sqrt{b(q-1)}%
}\right)  .
\]

The heat kernel%
\begin{equation}
K\left(  t\right)  =\frac{2^{\frac{1}{1-q}+\frac{3}{2}}\left[  b(q-1)/t^{2}%
\right]  ^{\frac{q-3}{4(q-1)}}}{\Gamma\left(  \frac{1}{q-1}-\frac{1}%
{2}\right)  }K_{\frac{1}{2}+\frac{1}{1-q}}\left(  \frac{t}{\sqrt{b(q-1)}%
}\right)  .
\end{equation}

The partition function%
\begin{equation}
Z\left(  \beta\right)  =\frac{2^{\frac{1}{1-q}+\frac{3}{2}}\left[
b(q-1)/\beta^{2}\right]  ^{\frac{q-3}{4(q-1)}}}{\Gamma\left(  \frac{1}%
{q-1}-\frac{1}{2}\right)  }K_{\frac{1}{2}+\frac{1}{1-q}}\left(  \frac{\beta
}{\sqrt{b(q-1)}}\right)  .
\end{equation}

The free energy%
\begin{equation}
F=-\frac{1}{\beta}\ln\frac{2^{\frac{1}{1-q}+\frac{3}{2}}\left[  \frac
{b(q-1)}{\beta^{2}}\right]  ^{\frac{q-3}{4(q-1)}}K_{\frac{1}{2}+\frac{1}{1-q}%
}\left(  \frac{\beta}{\sqrt{b(q-1)}}\right)  }{\Gamma\left(  \frac{1}%
{q-1}-\frac{1}{2}\right)  }.
\end{equation}

The internal energy%
\begin{equation}
U=\frac{K_{\frac{3}{2}+\frac{1}{1-q}}\left(  \frac{\beta}{\sqrt{b(q-1)}%
}\right)  +K_{\frac{q+1}{2-2q}}\left(  \frac{\beta}{\sqrt{b(q-1)}}\right)
}{2\sqrt{b(q-1)}K_{\frac{1}{2}+\frac{1}{1-q}}\left(  \frac{\beta}%
{\sqrt{b(q-1)}}\right)  }+\frac{q-3}{2\beta(q-1)}.
\end{equation}

The entropy%
\begin{align}
S\left(  \beta\right)   &  =\frac{1}{4\sqrt{b}(q-1)^{3/2}K_{\frac{1}{2}%
+\frac{1}{1-q}}\left(  \frac{\beta}{\sqrt{b(q-1)}}\right)  }\left\{
\sqrt{b(q-1)}K_{\frac{1}{2}+\frac{1}{1-q}}\left(  \frac{\beta}{\sqrt{b(q-1)}%
}\right)  \right. \nonumber\\
&  \times\left\{  4(q-1)\ln\left(  2^{\frac{1}{1-q}+\frac{3}{2}}\left[
\frac{b(q-1)}{\beta^{2}}\right]  ^{\frac{q-3}{4(q-1)}}K_{\frac{1}{2}+\frac
{1}{1-q}}\left(  \frac{\beta}{\sqrt{b(q-1)}}\right)  /\Gamma\left(  \frac
{1}{q-1}-\frac{1}{2}\right)  \right)  \right. \nonumber\\
&  +\left.  q-3\right\}  +2\beta(q-1)\left[  K_{\frac{3}{2}+\frac{1}{1-q}%
}\left(  \frac{\beta}{\sqrt{b(q-1)}}\right)  +K_{\frac{q+1}{2-2q}}\left(
\frac{\beta}{\sqrt{b(q-1)}}\right)  \right] \nonumber\\
&  +\left.  \beta(q-3)\sqrt{b(q-1)}K_{\frac{1}{2}+\frac{1}{1-q}}\left(
\frac{\beta}{\sqrt{b(q-1)}}\right)  \right\}  .
\end{align}

The specific heat%
\begin{align}
C_{V}  &  =\frac{q-3}{2(q-1)}+\frac{\beta^{2}}{2b(q-1)}+\frac{\beta^{2}\left[
K_{\frac{1}{1-q}-\frac{3}{2}}\left(  \frac{\beta}{\sqrt{b(q-1)}}\right)
+K_{\frac{5}{2}+\frac{1}{1-q}}\left(  \frac{\beta}{\sqrt{b(q-1)}}\right)
\right]  }{4b(q-1)K_{\frac{1}{2}+\frac{1}{1-q}}\left(  \frac{\beta}%
{\sqrt{b(q-1)}}\right)  }\nonumber\\
&  -\frac{\beta^{2}\left[  K_{\frac{3}{2}+\frac{1}{1-q}}\left(  \frac{\beta
}{\sqrt{b(q-1)}}\right)  +K_{\frac{q+1}{2-2q}}\left(  \frac{\beta}%
{\sqrt{b(q-1)}}\right)  \right]  ^{2}}{4b(q-1)K_{\frac{1}{2}+\frac{1}{1-q}%
}\left(  \frac{\beta}{\sqrt{b(q-1)}}\right)  ^{2}}.
\end{align}

\subsubsection{Quantum field}

The spectral zeta function%
\begin{equation}
\zeta\left(  s\right)  =\frac{\left[  1+(-1)^{s}\right]  \Gamma\left(
\frac{1}{2}-\frac{s}{2}\right)  \left[  b(q-1)\right]  ^{s/2}\Gamma\left(
\frac{s}{2}+\frac{1}{q-1}-\frac{1}{2}\right)  }{2\sqrt{\pi}\Gamma\left(
\frac{1}{q-1}-\frac{1}{2}\right)  }.
\end{equation}

When $s$ is a integer%
\begin{equation}
\zeta\left(  s\right)  =\frac{\left[  1+(-1)^{s}\right]  \Gamma\left(
\frac{1}{2}-\frac{s}{2}\right)  \left[  b(q-1)\right]  ^{s/2}\Gamma\left(
\frac{s}{2}+\frac{1}{q-1}-\frac{1}{2}\right)  }{2\sqrt{\pi}\Gamma\left(
\frac{1}{q-1}-\frac{1}{2}\right)  };
\end{equation}
when $s$ is even
\begin{equation}
\zeta\left(  s\right)  =\frac{\Gamma\left(  \frac{1}{2}-\frac{s}{2}\right)
\left[  b(q-1)\right]  ^{s/2}\Gamma\left(  \frac{s}{2}+\frac{1}{q-1}-\frac
{1}{2}\right)  }{\sqrt{\pi}\Gamma\left(  \frac{1}{q-1}-\frac{1}{2}\right)  };
\end{equation}
when $s$ is odd%
\begin{equation}
\zeta\left(  s\right)  =\frac{\left[  1+(-1)^{s}\right]  \Gamma\left(
\frac{1}{2}-\frac{s}{2}\right)  \left[  b(q-1)\right]  ^{s/2}\Gamma\left(
\frac{s}{2}+\frac{1}{q-1}-\frac{1}{2}\right)  }{2\sqrt{\pi}\Gamma\left(
\frac{1}{q-1}-\frac{1}{2}\right)  }=0.
\end{equation}

The one-loop effective action
\begin{equation}
W=\frac{1}{4}\left[  -\ln\left(  4b(q-1)\right)  -H_{\frac{1}{q-1}-\frac{3}%
{2}}+\pi\right]  ,
\end{equation}
where $H_{n}$ gives the $n$th harmonic number.

The vacuum energy%
\begin{equation}
E_{0}=0.
\end{equation}

\subsection{Intermediate distribution}

The intermediate distribution \cite{liu2012intermediate}%
\begin{equation}
p\left(  x,\mu,\sigma,\nu\right)  =\frac{1}{\nu\pi}e^{\left(  1/\nu-1\right)
\sigma^{2}}\int_{0}^{1}\cos\left(  \frac{x-\mu}{\nu}\ln t\right)
t^{\frac{\sigma^{2}}{\nu^{2}}-1}\exp\left(  \left(  1-\frac{1}{\nu^{2}%
}\right)  \sigma^{2}t\right)  dt.
\end{equation}
Working out the integral gives%
\begin{align}
p\left(  x,\mu,\sigma,\nu\right)   &  =\frac{1}{\nu\pi}\left(  \frac
{\sigma^{2}}{\nu^{2}}-\sigma^{2}\right)  ^{-\sigma^{2}/\nu^{2}}e^{\left(
1/\nu-1\right)  \sigma^{2}}\nonumber\\
&  \times\operatorname{Re}\left\{  \left(  \frac{\sigma^{2}}{\nu^{2}}%
-\sigma^{2}\right)  ^{i\left(  x-\mu\right)  /\nu}\left[  \Gamma\left(
\frac{\sigma^{2}}{\nu^{2}}-i\frac{x-\mu}{\nu}\right)  -\Gamma\left(
\frac{\sigma^{2}}{\nu^{2}}-i\frac{x-\mu}{\nu},\frac{\sigma^{2}}{\nu^{2}%
}-\sigma^{2}\right)  \right]  \right\}  .
\end{align}

\subsubsection{Thermodynamics}

The characteristic function%
\begin{equation}
f\left(  k\right)  =\exp\left(  ik\mu+\frac{\sigma^{2}}{\nu}\left[  \left(
e^{-\nu\left\vert k\right\vert }-1\right)  \left(  \nu-1\right)  -\left\vert
k\right\vert \right]  \right)  .
\end{equation}

The heat kernel%
\begin{equation}
K\left(  t\right)  =\exp\left(  -t\mu+\frac{\sigma^{2}}{\nu}\left[  \left(
e^{-\nu t}-1\right)  \left(  \nu-1\right)  -t\right]  \right)  .
\end{equation}

The partition function%
\begin{equation}
Z\left(  \beta\right)  =\exp\left(  -\beta\mu+\frac{\sigma^{2}}{\nu}\left[
\left(  e^{-\nu\beta}-1\right)  \left(  \nu-1\right)  -\beta\right]  \right)
.
\end{equation}

The free energy%
\begin{equation}
F=\mu+\frac{\sigma^{2}}{\nu}\left[  1-\frac{\nu-1}{\beta}\left(  e^{-\beta\nu
}-1\right)  \right]  .
\end{equation}

The internal energy%
\begin{equation}
U=\mu+\left[  e^{-\beta\nu}(\nu-1)+\frac{1}{\nu}\right]  \sigma^{2}.
\end{equation}

The entropy%
\begin{equation}
S=e^{-\beta\nu}(\nu-1)\beta\sigma^{2}+\frac{(\nu-1)\sigma^{2}}{\nu}\left(
e^{-\beta\nu}-1\right)  .
\end{equation}

The specific heat%
\begin{equation}
C_{V}=e^{-\beta\nu}\beta^{2}(\nu-1)\nu\sigma^{2}.
\end{equation}

\subsubsection{Quantum field}

The spectral zeta function%

\begin{equation}
\zeta\left(  s\right)  =\frac{1}{\pi}\nu^{-s}\sin(\pi s)\Gamma(1-s)\exp\left(
-\frac{i\pi s}{2}+\frac{\sigma^{2}\left(  1-\nu\right)  }{\nu}\right)
\int_{0}^{1}\exp\left(  t\frac{\sigma^{2}\left(  \nu^{2}-1\right)  }{\nu^{2}%
}\right)  t^{\frac{i\mu}{\nu}+\frac{\sigma^{2}}{\nu^{2}}-1}\left(  \ln
t\right)  ^{s-1}dt.
\end{equation}

The one-loop effective action
\begin{equation}
W=\frac{1}{2}e^{-\sigma^{2}\left(  1-1/\nu\right)  }\int_{0}^{1}\exp\left(
t\frac{\sigma^{2}\left(  \nu-1\right)  }{\nu}+\left(  \frac{i\mu}{\nu}%
+\frac{\sigma^{2}}{\nu^{2}}-1\right)  \ln t\right)  \frac{1}{\ln t}dt.
\end{equation}

The vacuum energy%
\begin{equation}
E_{0}=\mu-i\sigma^{2}\left(  1-\frac{1}{\nu}-\nu\right)  .
\end{equation}

\section{Intermediate-distribution distribution function
\label{Inter}}

In Ref. \cite{liu2012intermediate}, we proposed an intermediate distribution
between the Cauchy distribution and the Gaussian distribution. The Cauchy and
Gaussian distributions are two special cases of the intermediate distribution,
respectively. The Cauchy and Gaussian distributions are two very different
distributions, as evidenced by the fact that the Cauchy distribution lacks
moments while the Gaussian distribution has moments
\cite{zhang2022renormalization}. In this appendix, we give an alternative
derivation of the distribution function for the intermediate distribution by
directly working out the integral in the distribution function of the
intermediate distribution.

The probability density function of the intermediate distribution is
\cite{liu2012intermediate}%
\begin{equation}
p\left(  x,\mu,\sigma,\nu\right)  =\frac{1}{\nu\pi}e^{\left(  1/\nu-1\right)
\sigma^{2}}\int_{0}^{1}\cos\left(  \frac{x-\mu}{\nu}\ln t\right)
t^{\frac{\sigma^{2}}{\nu^{2}}-1}e^{\left(  1-1/\nu^{2}\right)  \sigma^{2}t}dt.
\label{PDF}%
\end{equation}

Introduce the integral%
\begin{equation}
I\left(  \tau\right)  =\int_{0}^{\tau}\cos\left(  \frac{x-\mu}{\nu}\ln
t\right)  t^{\frac{\sigma^{2}}{\nu^{2}}-1}e^{\left(  1-1/\nu^{2}\right)
\sigma^{2}t}dt, \label{Itau}%
\end{equation}
and we have $p\left(  x,\mu,\sigma,\nu\right)  =\frac{1}{\nu\pi}e^{\left(
1/\nu-1\right)  \sigma^{2}}I\left(  1\right)  $.

Rewrite the integral (\ref{Itau}) as%
\begin{align}
I\left(  \tau\right)   &  =e^{\left(  1-1/\nu^{2}\right)  \sigma^{2}\tau}%
\int_{0}^{\tau}\cos\left(  \frac{x-\mu}{\nu}\ln t\right)  t^{\frac{\sigma^{2}%
}{\nu^{2}}-1}e^{\left(  1/\nu^{2}-1\right)  \sigma^{2}\tau}e^{\left(
1-1/\nu^{2}\right)  \sigma^{2}t}dt\nonumber\\
&  =e^{\left(  1-1/\nu^{2}\right)  \sigma^{2}\tau}\mathcal{I}\left(
\tau\right)  . \label{IandI}%
\end{align}
The integral $\mathcal{I}\left(  \tau\right)  $ can be expressed using the
Laplace-Type convolution:%
\begin{align}
\mathcal{I}\left(  \tau\right)   &  =\int_{0}^{\tau}\cos\left(  \frac{x-\mu
}{\nu}\ln t\right)  t^{\frac{\sigma^{2}}{\nu^{2}}-1}\exp\left(  \left(
\frac{1}{\nu^{2}}-1\right)  \sigma^{2}\left(  \tau-t\right)  \right)
dt\nonumber\\
&  =\cos\left(  \frac{x-\mu}{\nu}\ln\tau\right)  \tau^{\frac{\sigma^{2}}%
{\nu^{2}}-1}\ast\exp\left(  \left(  \frac{1}{\nu^{2}}-1\right)  \sigma^{2}%
\tau\right)  dt, \label{scItau}%
\end{align}
where $\ast$ denotes the convolution.

Using the convolution theorem of the Laplace transform
\cite{poularikas2018transforms}%
\begin{equation}
\mathcal{L}\left[  f\left(  t\right)  \ast g\left(  t\right)  \right]
=\mathcal{L}\left[  f\left(  t\right)  \right]  \mathcal{L}\left[  g\left(
t\right)  \right]  ,
\end{equation}
we can rewrite the integral (\ref{scItau}) as%
\begin{equation}
\mathcal{I}\left(  \tau\right)  =\mathcal{L}^{-1}\left(  \mathcal{L}\left[
\cos\left(  \frac{x-\mu}{\nu}\ln\tau\right)  \tau^{\frac{\sigma^{2}}{\nu^{2}%
}-1}\right]  \mathcal{L}\left[  \exp\left(  \left(  \frac{1}{\nu^{2}%
}-1\right)  \sigma^{2}\tau\right)  \right]  \right)  . \label{LLL}%
\end{equation}
Working out the two Laplace transforms in Eq. (\ref{LLL}),%
\begin{align}
\mathcal{L}\left[  \cos\left(  \frac{x-\mu}{\nu}\ln\tau\right)  \tau
^{\frac{\sigma^{2}}{\nu^{2}}-1}\right]   &  =\frac{1}{2}s^{-i\frac{x-\mu}{\nu
}-\frac{\sigma^{2}}{\nu^{2}}}\left[  s^{2i\frac{x-\mu}{\nu}}\Gamma\left(
-i\frac{x-\mu}{\nu}+\frac{\sigma^{2}}{\nu^{2}}\right)  +\Gamma\left(
i\frac{x-\mu}{\nu}+\frac{\sigma^{2}}{\nu^{2}}\right)  \right]  ,\\
\mathcal{L}\left[  \exp\left(  \left(  \frac{1}{\nu^{2}}-1\right)  \sigma
^{2}\tau\right)  \right]   &  =\frac{1}{s-\left(  \frac{1}{\nu^{2}}-1\right)
\sigma^{2}},
\end{align}
gives%
\begin{align}
\mathcal{I}\left(  \tau\right)   &  =\mathcal{L}^{-1}\left(  \frac{1}%
{2}s^{-i\frac{x-\mu}{\nu}-\frac{\sigma^{2}}{\nu^{2}}}\left(  s^{2i\frac{x-\mu
}{\nu}}\Gamma\left(  -i\frac{x-\mu}{\nu}+\frac{\sigma^{2}}{\nu^{2}}\right)
+\Gamma\left(  i\frac{x-\mu}{\nu}+\frac{\sigma^{2}}{\nu^{2}}\right)  \right)
\frac{1}{s-\left(  \frac{1}{\nu^{2}}-1\right)  \sigma^{2}}\right) \nonumber\\
&  =\frac{1}{2}e^{\left(  \frac{1}{\nu^{2}}-1\right)  \sigma^{2}\tau}\left(
\frac{\sigma^{2}}{\nu^{2}}-\sigma^{2}\right)  ^{-i\frac{x-\mu}{\nu}%
-\frac{\sigma^{2}}{\nu^{2}}}\nonumber\\
&  \times\left\{  \left(  \frac{\sigma^{2}}{\nu^{2}}-\sigma^{2}\right)
^{2i\frac{x-\mu}{\nu}}\left[  \Gamma\left(  -i\frac{x-\mu}{\nu}+\frac
{\sigma^{2}}{\nu^{2}}\right)  -\Gamma\left(  -i\frac{x-\mu}{\nu}+\frac
{\sigma^{2}}{\nu^{2}},\left(  \frac{1}{\nu^{2}}-1\right)  \sigma^{2}%
\tau\right)  \right]  \right. \nonumber\\
&  \left.  +\Gamma\left(  i\frac{x-\mu}{\nu}+\frac{\sigma^{2}}{\nu^{2}%
}\right)  -\Gamma\left(  i\frac{x-\mu}{\nu}+\frac{\sigma^{2}}{\nu^{2}},\left(
\frac{1}{\nu^{2}}-1\right)  \sigma^{2}\tau\right)  \right\}  , \label{scI2}%
\end{align}
where $\Gamma\left(  \alpha\right)  $ is the gamma function and $\Gamma\left(
\alpha,\beta\right)  $ is the incomplete gamma function \cite{olver2010nist}.

Substituting Eq. (\ref{scI2}) into Eq. (\ref{IandI}) and taking $\tau=1$ give
an explicit expression of the intermediate-distribution distribution
function,
\begin{align}
p\left(  x,\mu,\sigma,\nu\right)   &  =\frac{1}{\nu\pi}\left(  \frac
{\sigma^{2}}{\nu^{2}}-\sigma^{2}\right)  ^{-\sigma^{2}/\nu^{2}}e^{\left(
1/\nu-1\right)  \sigma^{2}}\nonumber\\
&  \times\operatorname{Re}\left\{  \left(  \frac{\sigma^{2}}{\nu^{2}}%
-\sigma^{2}\right)  ^{i\left(  x-\mu\right)  /\nu}\left[  \Gamma\left(
\frac{\sigma^{2}}{\nu^{2}}-i\frac{x-\mu}{\nu}\right)  -\Gamma\left(
\frac{\sigma^{2}}{\nu^{2}}-i\frac{x-\mu}{\nu},\frac{\sigma^{2}}{\nu^{2}%
}-\sigma^{2}\right)  \right]  \right\}  .
\end{align}
\bigskip
\bigskip
\bigskip

%\hrule

%\appendix

\acknowledgments

We are very indebted to Dr G. Zeitrauman for his encouragement. This work is supported in part by Special Funds for theoretical physics Research Program of the NSFC under Grant No.
11947124, and NSFC under Grant Nos. 11575125 and 11675119.

\nolinenumbers
%%%%%%%%%%%%%%%%%%%%%%%%%%%%%%%%%%%%%%%%%%

%%%%%%%%%%%%%%%%%%%%%%%%%%%%%%%%%%%%%%%%%%
% To add notes in main text, please use \endnote{} and un-comment the codes below.
%\begin{adjustwidth}{-5.0cm}{0cm}
%\printendnotes[custom]
%\end{adjustwidth}
%%%%%%%%%%%%%%%%%%%%%%%%%%%%%%%%%%%%%%%%%%
\reftitle{References}


%\externalbibliography{yes}
%\bibliography{your_external_BibTeX_file}

%=====================================
% References, variant B: internal bibliography
%=====================================

\providecommand{\href}[2]{#2}\begingroup\raggedright\endgroup

%\bibliographystyle{JHEP} %²Î¿¼ÎÄÏ×µÄ·ç¸ñ(.bst)
%\bibliography{refs} %²Î¿¼ÎÄÏ×ÎÄ¼þ(.bib)
\end{paracol}
% If authors have biography, please use the format below
%\section*{Short Biography of Authors}
%\bio
%{\raisebox{-0.35cm}{\includegraphics[width=3.5cm,height=5.3cm,clip,keepaspectratio]{Definitions/author1.pdf}}}
%{\textbf{Firstname Lastname} Biography of first author}
%
%\bio
%{\raisebox{-0.35cm}{\includegraphics[width=3.5cm,height=5.3cm,clip,keepaspectratio]{Definitions/author2.jpg}}}
%{\textbf{Firstname Lastname} Biography of second author}

% The following MDPI journals use author-date citation: Admsci,  Arts, Econometrics, Economies, Genealogy, Humanities, IJFS, Jintelligence, JRFM, Languages, Laws, Literature, Religions, Risks, Social Sciences. For those journals, please follow the formatting guidelines on http://www.mdpi.com/authors/references
% To cite two works by the same author: \citeauthor{ref-journal-1a} (\citeyear{ref-journal-1a}, \citeyear{ref-journal-1b}). This produces: Whittaker (1967, 1975)
% To cite two works by the same author with specific pages: \citeauthor{ref-journal-3a} (\citeyear{ref-journal-3a}, p. 328; \citeyear{ref-journal-3b}, p.475). This produces: Wong (1999, p. 328; 2000, p. 475)

%%%%%%%%%%%%%%%%%%%%%%%%%%%%%%%%%%%%%%%%%%
%% for journal Sci
%\reviewreports{\\
%Reviewer 1 comments and authors’ response\\
%Reviewer 2 comments and authors’ response\\
%Reviewer 3 comments and authors’ response
%}
%%%%%%%%%%%%%%%%%%%%%%%%%%%%%%%%%%%%%%%%%%
\end{document}